# Effect of Ag nano-additivation on microstructure formation in Nd-Fe-B magnets built by laser powder bed fusion

Varatharaja Nallathambi [a,b], Philipp Gabriel [a], Xinren Chen [b], Ziyuan Rao [b], Konstantin Skokov [c], Oliver Gutfleisch [c], Stephan Barcikowski [a], Anna Rosa Ziefuss [a]*, Baptiste Gault [b,d]*

[a] Technical Chemistry I and Center for Nanointegration Duisburg-Essen (CENIDE), University of Duisburg-Essen, 45141 Essen, Germany

[b] Max Planck Institute for Sustainable Materials, 40237 Düsseldorf, Germany

[c] Functional Materials, Institute of Material Science, Technical University of Darmstadt, 64287 Darmstadt, Germany

[d] Department of Materials, Royal School of Mines, Imperial College London, London SW72AZ, UK

* corresponding authors: b.gault@mpie.de, anna.ziefuss@uni-due.de

Keywords: Permanent magnets, Additive manufacturing, Surface additivation, Grain refinement

## Abstract

Laser powder bed fusion (PBF-LB/M) enables the near-net shape production of permanent magnets with complex geometry while reducing material waste. However, controlling the microstructure and optimizing magnetic properties remain challenging due to rapid solidification and intrinsic heat treatment effects occurring during both inter-layer and intra-layer processing. Surface additivation of the feedstock powder with Ag nanoparticles (NPs) is a concept that has been shown to increase the coercivity of PBF-LB/M-produced Nd-Fe-B magnets. Using atom probe tomography (APT) and transmission electron microscopy (TEM), we reveal that Ag nano-additivation promotes heterogeneous nucleation of the $Nd_2Fe_{14}B$ phase, leading to refined, equiaxed grains and increased stability of the Ti-Zr-B-rich intergranular phase. The intrinsic heat treatment, influenced by layer-wise processing, further affects the distribution of Ag-rich regions, impacting grain growth and intergranular phase composition across different regions of the melt pool. Compared to the unadditivated sample, the Ag-additivated sample exhibits a significantly finer grain structure and a changed intergranular phase, which contribute to enhanced domain wall pinning and coercivity. These microstructural changes directly modify the magnetic domain structure, as evidenced by Lorentz transmission electron microscopy (TEM). Our results highlight that the interplay between nano-additivation and in-process heat treatment provides a novel pathway for tailoring the microstructure and enhancing the magnetic performance of permanent magnets.

## 1. Introduction

The transition towards clean energy and sustainable practices creates a much-increased demand for new permanent magnets for electric vehicles, power generators, wind turbines, and consumer electronics [1,2]. Rare earth (RE)-based permanent magnets, particularly Nd-Fe-B, are the strongest and most commonly used option for applications at room temperature [1]. The demand for RE-based magnets is predicted to increase significantly by roughly 3 times by 2030 despite the criticality of the supply of RE elements [3]. In addition, the production of commercial sintered Nd-Fe-B magnets leads to the loss of valuable RE elements in the form of swarf, which are the waste powders and filings that result from manufacturing. The wastage amounts up to 73% by mass of the materials, depending on the cutting and machining processes involved [4–6]. This has motivated a search for alternative manufacturing strategies to reduce RE wastage, along with research in the recovery and recycling of the RE from magnet waste, including swarf [7–10].

Additive manufacturing can produce near-net-shaped objects, enabling improved material usage [11–13]. Recent research has shown promising potential in laser powder bed fusion of metals (PBF-LB/M) to manufacture shaped Nd-Fe-B-based permanent magnets [14–17]. Beyond geometrical freedom of shape and achieving dense metallic parts potentially with reasonable surface finish, PBF-LB/M offers opportunities for microstructure control through adjustments of the process parameters (e.g. energy input and scan strategy,) or the use of nano-sized surface additives in the feedstock powder to optimize structural and functional properties [18,19]. As such, PBF-LB/M holds prospects for enhancing magnetic performance by manipulating the microstructure and composition of the phases formed. The microstructure of the conventional sintered Nd-Fe-B-based permanent magnets consists primarily of the ferromagnetic tetragonal $Nd_2Fe_{14}B$ (2:14:1) grains surrounded by paramagnetic Nd-rich grain boundary regions, also referred to as the intergranular phase and other precipitate phases at the grain boundaries or grain junctions [20].

In sintered Nd-Fe-B magnets, previous investigations have demonstrated the influence of the composition and magnetic properties of the thin intergranular phase region that surrounds the primary $Nd_2Fe_{14}B$ phase [20–28]. The Nd-rich intergranular phase helps in decoupling the magnetic $Nd_2Fe_{14}B$ grains, which increases the coercivity [29,30]. However, the commercially available Nd-Fe-B feedstock powder, gas-atomized MQP-S-11-9-20001 ($Nd_{7.5}Pr_{0.7}Zr_{2.6}Ti_{2.5}Co_{2.5}Fe_{75.4}B_{8.8}$ at. %) [31] is Nd-lean (as the Nd concentration is less than the minimum 11.8 at.% required to form the stoichiometric line compound $Nd_2Fe_{14}B$) compared to the powders used for conventional RE-excess sintered magnets that contain 14–18 at.% Nd [20,21]. As a result, the comparably higher Fe content leads to the formation of a magnetically soft α-Fe phase with high magnetization, which is exchange-coupled with the magnetically hard $Nd_2Fe_{14}B$ phase, which decreases coercivity to some extent but leads to enhancement in isotropic remanence and squareness of the hysteresis loop [31]. As MQP-S is readily available commercially and has a spherical morphology, it serves as a good choice for PBF-LB/M processing. The intergranular phase formed in this case is Nd-depleted [32] with substantial amounts of Fe and B and alloying additions like Ti and Zr that also support the grain refinement of the 2:14:1 matrix phase.

Surface additivation of the feedstock powder with nanoparticles (NPs) provides opportunities for compositional modification during the PBF-LB/M processing as it enables melting and mixing of the NPs with the liquid melt [19]. In our previous work [33], we recently reported that the addition of 1 wt.% Ag NPs enhances the coercivity of PBF-LB/M-built Nd-Fe-B magnets to a record value of 935 ± 6 kA $m^{-1}$ which is 17 ± 6% higher than the unadditivated counterpart. The increase in coercivity was achieved without compromising on the remanence and further improving the part density by around 10% without requiring any heat treatment or post-processing. This increase in coercivity with Ag nano-additivation was mainly attributed to the grain refinement of the melt pool microstructure after PBF-LB/M processing [33]. However, further investigation was required to better understand the mechanisms whereby Ag nano-additivation leads to a modification of the microstructure and phase composition that underpin the enhanced properties. In particular, the role of the intrinsic heat treatment, which inevitably occurs during layer-wise PBF-LB/M processing, has not been systematically considered in previous studies. Understanding how Ag nano-additivation interacts with this inherent thermal cycling is crucial for unraveling the full impact on microstructure evolution and magnetic performance.

Here, we report on a study of the microstructure of the unadditivated and Ag-additivated as-built parts using the inherent complementarities of transmission electron microscopy (TEM) and atom probe tomography (APT) that allows us to reveal 3D compositional and morphological differences. Based on these observations, we propose a plausible explanation of the distinct microstructure formation between the two as-built parts. Fresnel-mode Lorentz TEM imaging reveals the magnetic domain structure differences between the two as-built Nd-Fe-B magnets, evidencing the intricate relationship between the microstructure of the melt pool and the intergranular phase composition. These, in combination, underpin the observed property enhancement from Ag NPs surface additivation.

## 2. Materials and Methods

Details of the sample fabrication can be found in Gabriel and Nallathambi et al. [33]. In short, Nd-Fe-B permanent magnet parts were built by using PBF-LB/M (Trumpf – TruPrint 1000) from the MQP-S-11-9-20001 commercial feedstock powder (referred to as unadditivated) and from the same powder modified with approx. 10-nm-diameter laser-generated Ag NPs (referred to as Ag-additivated). The applied process parameters, laser power of 74 W, scan speed of 2300 mm $s^{-1}$, hatch distance of 15 μm, and layer thickness of 30 μm, leading to a volumetric energy density (VED) of 71.5 J $mm^{-3}$, were optimized to maximize the coercivity [33].

A dual beam scanning-electron microscope - Ga-ion focused ion beam (SEM-FIB) microscope (Thermo Fischer Helios 600i) was used to prepare the TEM specimens and a series of needle-shaped APT specimens by in-situ lift-out method [34] in the area of interest, followed by annular milling until the tip radius is less than 100 nm (**Figure S1**). APT measurements were carried out using a Cameca LEAP 5000 XR, with a base temperature of 50 K, pulsed UV-laser mode at an energy of 45 to 65 pJ per pulse, a target detection rate of 0.5 % and a repetition rate of 125 kHz. Data reconstruction and post-processing were done using the software AP Suite by Cameca Instruments following the voltage-based reconstruction protocol (a representative mass spectrum can be found in **Figure S2**).

The microstructure and elemental distribution in different phases of the as-built parts were studied using scanning transmission electron microscopy (STEM) imaging combined with energy dispersive X-ray spectroscopy (EDS) (Thermo Fischer Titan Themis 300 – probe corrected). The microscope was operated in STEM mode at an accelerating voltage of 300 kV, camera length of 100 mm, beam convergence angle of 23.8 mrad and a high-angle annular dark field (HAADF) detector was used with a collection angle range of 78 – 200 mrad.

Lorentz transmission electron microscopy (TEM) analysis was performed using an image-corrected Thermo Fischer Titan Themis G2 microscope operated at 300 kV. The TEM specimens were demagnetized before the analysis. Fresnel-mode Lorentz images, both in-focus and out-of-focus, were obtained under magnetic field-free conditions by fully deactivating the objective lens. Subsequently, in-situ experiments were conducted by activating and adjusting the objective lens to 80% of its full strength after switching the microscope to normal TEM mode. Due to interference from other lenses, the magnetic field generated at 80% objective lens intensity ranged between 1000 and 2520 kA $m^{-1}$. The microscope was then switched back to Lorentz mode, and changes in the domain walls were observed using Fresnel-mode imaging.

## 3. Results and Discussion

The effects of the Ag nano-additivation were studied by comparing the microstructure formation, phase composition (matrix, intergranular and precipitate phases), and process-inherent heat effects of the unadditivated and Ag-additivated as-built PBF-LB/M parts followed by their relations to the observed magnetic properties. **Figure 1** schematically illustrates the impact of the layer-wise production technique during the PBF-LB/M processing on the formation of the microstructure, with successions of melt pools separated by melt pool boundary regions. This schematic is used in the following sections to indicate the extraction locations of the specimen for TEM and APT characterization.

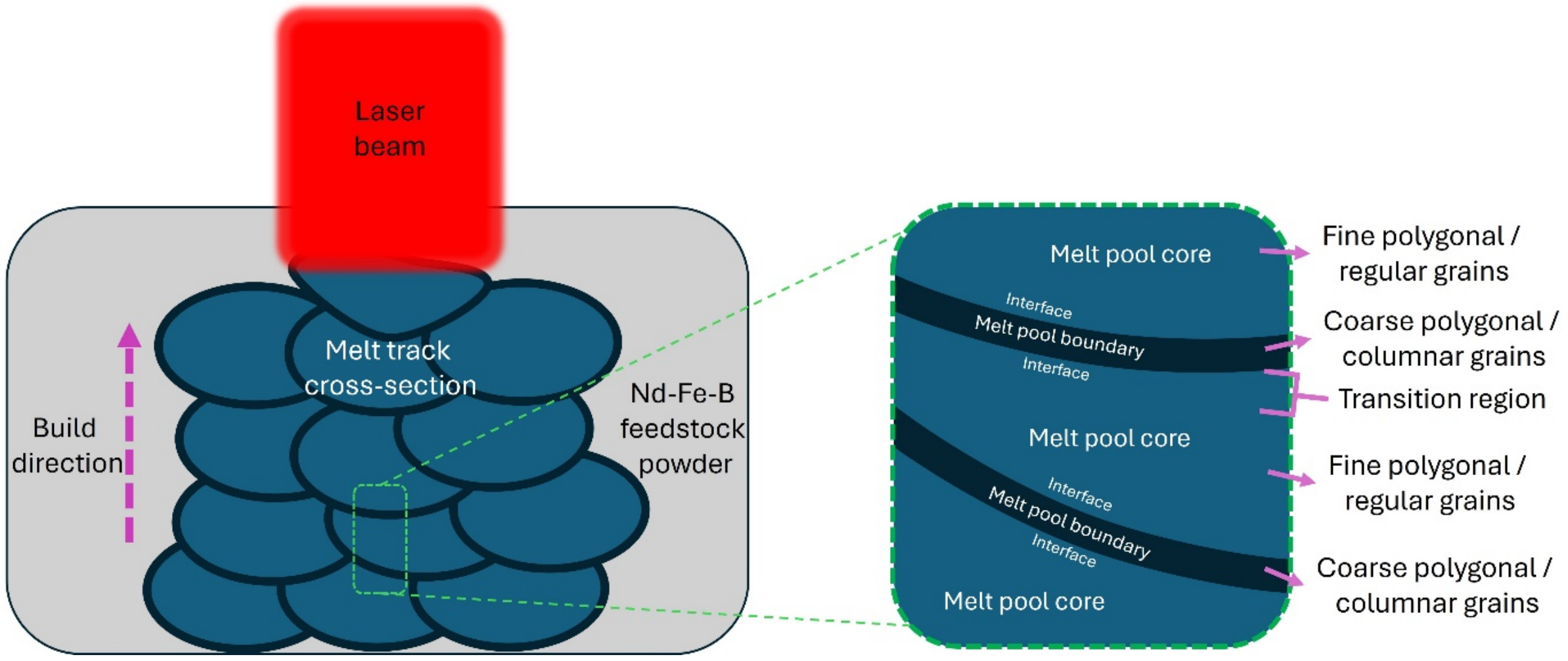

*Figure 1 Schematic illustration of melt track cross-section (left) and representative melt pool regions with expected grain morphology (right) formed along the build direction during the PBF-LB/M processing.*

### 3.1 Microstructure and phase composition analysis

The additively manufactured Nd-Fe-B-based magnets exhibit a microstructure comprising melt pools aligning one above the other across the build direction (**Figure 1**), with Ag nano-additivation altering the grain size distribution [33]. STEM high-angle annular dark field (HAADF) images shown in **Figure 2** highlight the grain size distribution and morphology differences observed between the unadditivated (**Figures 2a and b**) and Ag-additivated samples (**Figures 2c and d**) in the melt pool core. The melt pool core of the unadditivated sample consists of elongated irregular polygonal grains with a diameter ranging from 80 nm to as large as 350 nm, while the Ag-modified sample consists of a more equiaxed grain structure with finer grains as small as 30 nm in the melt pool core. Overall, the unadditivated sample shows a wider grain size distribution with an average grain size of 160 ± 67 nm (diameter equivalent) and the Ag-additivated sample with an average grain size of 58 ± 19 nm (**Figure S3**). The grain size and morphology differences observed from an irregular polygonal shape in the unadditivated sample to a regular cubic-shaped grain structure in the Ag-additivated sample indicate the different microstructure formation steps involved during the solidification of the melt pool. The primary $Nd_2Fe_{14}B$ grains appearing bright (Nd-Pr-rich) can be seen separated by the thin intergranular phase appearing dark (Ti-Zr-rich & Nd-Pr-lean) in the STEM-HAADF images. The elemental distribution in different phases of the Ag-additivated sample from STEM-EDS analysis is shown in **Figure 2e** and for the unadditivated sample in **Figure S4**. Spots of Nd-rich precipitate regions can be observed in both samples. Additionally, in the Ag-additivated sample, nano-sized Ag-rich regions can be seen distributed along the intergranular phase region (**Figure 2e**). The microstructural features observed align well with the findings reported earlier [33]. To further gain insights into the three-dimensional (3D) element distribution in different phases found in the microstructure, APT analyses of specimens prepared from the melt pool core of the unadditivated and Ag-additivated samples were performed.

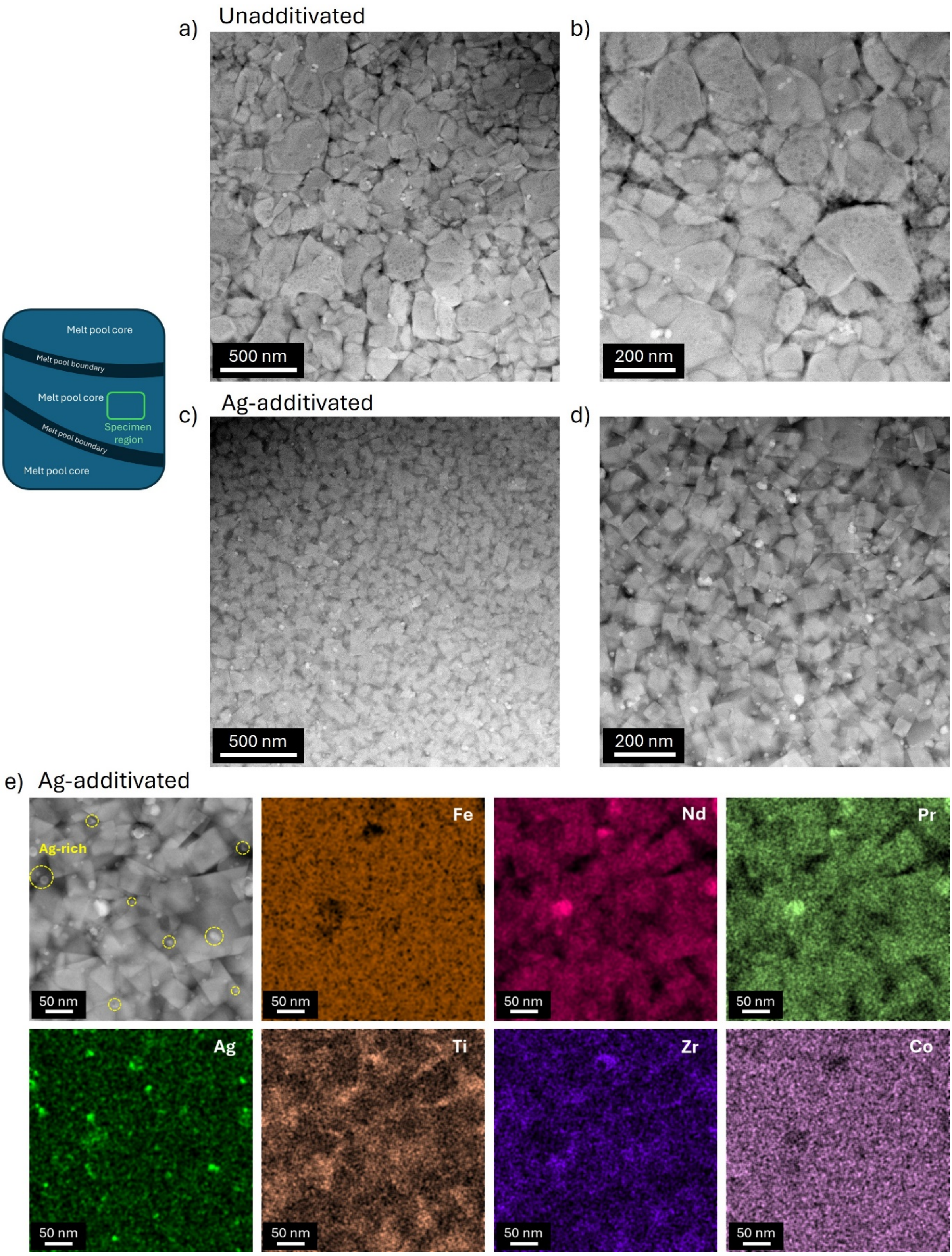

*Figure 2 STEM characterization of the melt pool core region of as-built parts. a), b) STEM-HAADF micrographs of the unadditivated and c), d) Ag-additivated samples showing the grain size distribution and morphology differences between them. e) STEM-EDS mappings of the Ag-additivated sample. Exemplary Ag-rich regions are highlighted in dotted yellow circles in the STEM-HAADF image.*

The 3D reconstructed APT map of the unadditivated sample is shown in **Figure 3a**. A predominantly nanocrystalline grain structure likely resulting from the fast cooling of the melt pool during the PBF-LB/M process is observed. The primary $Nd_2Fe_{14}B$ phase (Nd-Pr-rich) is compositionally distinct from the intergranular phase (Ti-Zr-B-rich) and the microstructure comprising thin intergranular phase regions surrounding the $Nd_2Fe_{14}B$ grains is visualized by constructing a Ti isosurface (grey) with a concentration greater than 3.5 at.%, as shown in **Figure 3b**.

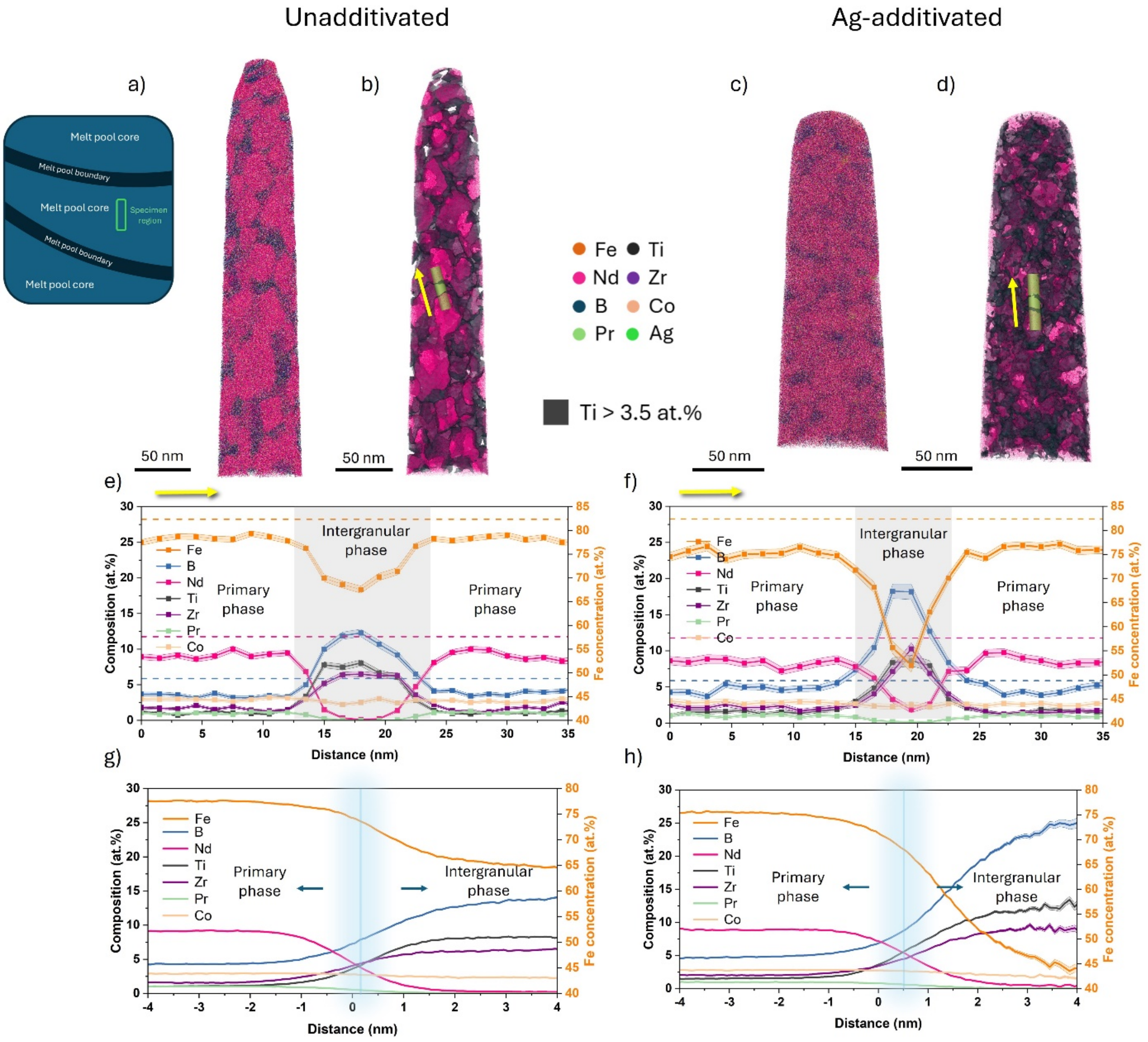

*Figure 3 APT results comparison of the unadditivated and Ag-additivated samples. a), c) are the 3D reconstructed atom maps; b), d) 3D atom maps highlighting the Ti-Zr-B-rich intergranular phase with an isosurface of Ti (grey) at a concentration greater than 3.5 at.%; e), f) 1D composition profile of individual elements using cylindrical ROIs shown in (b) and (d) (yellow arrows) across the primary and intergranular phases; g), h) Composition proxigrams constructed with Ti isosurface (grey) of concentration greater than 3.5 at.%, highlighting the distribution of individual elements between the primary and intergranular phases of the unadditivated and Ag-additivated samples, respectively. The*

*concentration of Fe is plotted on the right Y-axis for better visualization. Dashed lines representing nominal compositions for Fe (orange), Nd (pink) and B (blue) of a stoichiometric $Nd_2Fe_{14}B$ phase are shown as guide marks in e) and f). Alloying additions of Pr, Co, Zr and Ti compensate for the rest.*

**Figures 3c and d** show the 3D reconstructed APT map of the Ag-additivated sample and the corresponding Ti isosurface (grey) construction with a concentration greater than 3.5 at.% highlighting the intergranular phase regions, respectively. The set of Ti isosurfaces in grey confirms the findings of the TEM analysis, which reveal finer, more regular grains in the Ag-additivated sample (**Figure 3d**) compared to irregular polygonal grains of the unadditivated sample (**Figure 3b**) which helps in achieving overall better magnetic properties [35–38]. One-dimensional (1D) composition profiles calculated along cylindrical regions of interest (ROI) positioned across the two phases are plotted in **Figures 3e and f**. The primary $Nd_2Fe_{14}B$ phase in both samples contains around 9 at.% Nd with approx. 4 at.% B, 3 at.% Co and 1–2 at.% of Pr, Ti, and Zr and the remaining Fe with no noticeable amounts of Ag in the Ag-additivated sample. The intergranular phase is enriched with Ti, Zr and B with a depletion of RE elements Nd and Pr agreeing with previous reports for a Nd-lean feedstock [32]. Co is equally distributed in both phases.

### 3.2 Differences in the intergranular phase region

The set of Ti isosurfaces constructed (**Figures 3b and d**) shows that the intergranular phase is slightly thinner in the Ag-additivated sample, typically 5 ± 1 nm full-width half maximum (FWHM), compared to its unadditivated counterpart where it is closer to 10 ± 1 nm (highlighted in grey in **Figures 3e and f**). Composition profiles as a function of the distance to the Ti isosurfaces displayed in **Figures 3b and d**, i.e. proximity histograms or proxigrams [39], were used to determine the average compositions of the two phases. These are plotted in **Figures 3g and h** for the unadditivated and the Ag-additivated samples, respectively, and the average concentration values are tabulated in **Table 1**. The compositions in the primary phase are comparable, with a slightly higher Fe content (77.6 at.%) in the unadditivated sample than in the Ag-additivated sample (75.6 at.%). Compared to the unadditivated sample, the concentrations of Ti, Zr and B in the intergranular phase of the Ag-additivated sample are much higher (resp. 6.2, 8.3 and 13.7 at.% vs. 9.6, 12.1 and 22.9 at.%), whereas the Fe content drops from 65 at.% to an average value of 46 at.%. The presence of low amounts of RE elements in the intergranular phase (0.56 at.% Nd and 0.06 at.% Pr) ensures their effective utilization for the formation of the $Nd_2Fe_{14}B$ hard-magnetic phase after the PBF-LB/M manufacturing process. From the compositions of the different phases, the lever rule was used to determine the phase volume fractions [40,41]. The relative volume fraction of the intergranular phase in the unadditivated sample amounts to 19.7 ± 1.2 % compared to only 9.7 ± 4.2 % in the Ag-additivated sample (**Figure S5**). This translates to a higher volume fraction of the $Nd_2Fe_{14}B$ hard magnetic phase in the Ag-additivated sample which contributes to enhancing the coercivity.

*Table 1 Comparison of average elemental concentrations in the primary and the intergranular phases of the unadditivated and Ag-additivated samples from APT.*

| Elements (at. %) | Fe | B | Nd | Ti | Zr | Pr | Co |
|---|---|---|---|---|---|---|---|
| Primary $Nd_2Fe_{14}B$ phase | | | | | | | |
| Unadditivated sample | 77.62 ± 0.06 | 4.30 ± 0.03 | 9.15 ± 0.04 | 1.07 ± 0.02 | 1.55 ± 0.02 | 1.04 ± 0.02 | 2.87 ± 0.03 |
| Ag-additivated sample | 75.61 ± 0.07 | 4.79 ± 0.04 | 8.89 ± 0.05 | 1.53 ± 0.02 | 2.01 ± 0.02 | 0.98 ± 0.02 | 2.73 ± 0.03 |
| Intergranular phase | | | | | | | |
| Unadditivated sample | 65.17 ± 0.13 | 13.72 ± 0.1 | 0.28 ± 0.01 | 8.33 ± 0.08 | 6.20 ± 0.07 | 0.04 ± 0.01 | 2.33 ± 0.04 |
| Ag-additivated sample | 46.44 ± 0.38 | 22.94 ± 0.32 | 0.56 ± 0.06 | 12.08 ± 0.25 | 9.65 ± 0.22 | 0.06 ± 0.02 | 1.73 ± 0.1 |

Trace amounts of Cu, C, Cr, Al, P, O, and Si were in the remaining.

To support our interpretation, an APT dataset from a partially melted/spattered Nd-Fe-B feedstock powder found on the surface of the as-built Ag-additivated part is shown in **Figures 4a–c**. It shows a large $Nd_2Fe_{14}B$ grain sandwiched between the intergranular phase, providing a glimpse of the grain structure of the Nd-Fe-B feedstock micropowder. This confirms that the nanocrystalline microstructure is formed through the melting of the feedstock micropowder and over the intrinsic heat treatment during the PBF-LB/M processing. Another set of representative 3D reconstructions can be seen in **Figures 4d–g** for the two as-built permanent magnets prepared from the melt pool core, highlighting the grain structure distribution and morphology differences between them which is also evidenced in STEM-HAADF images shown in **Figure 2**.

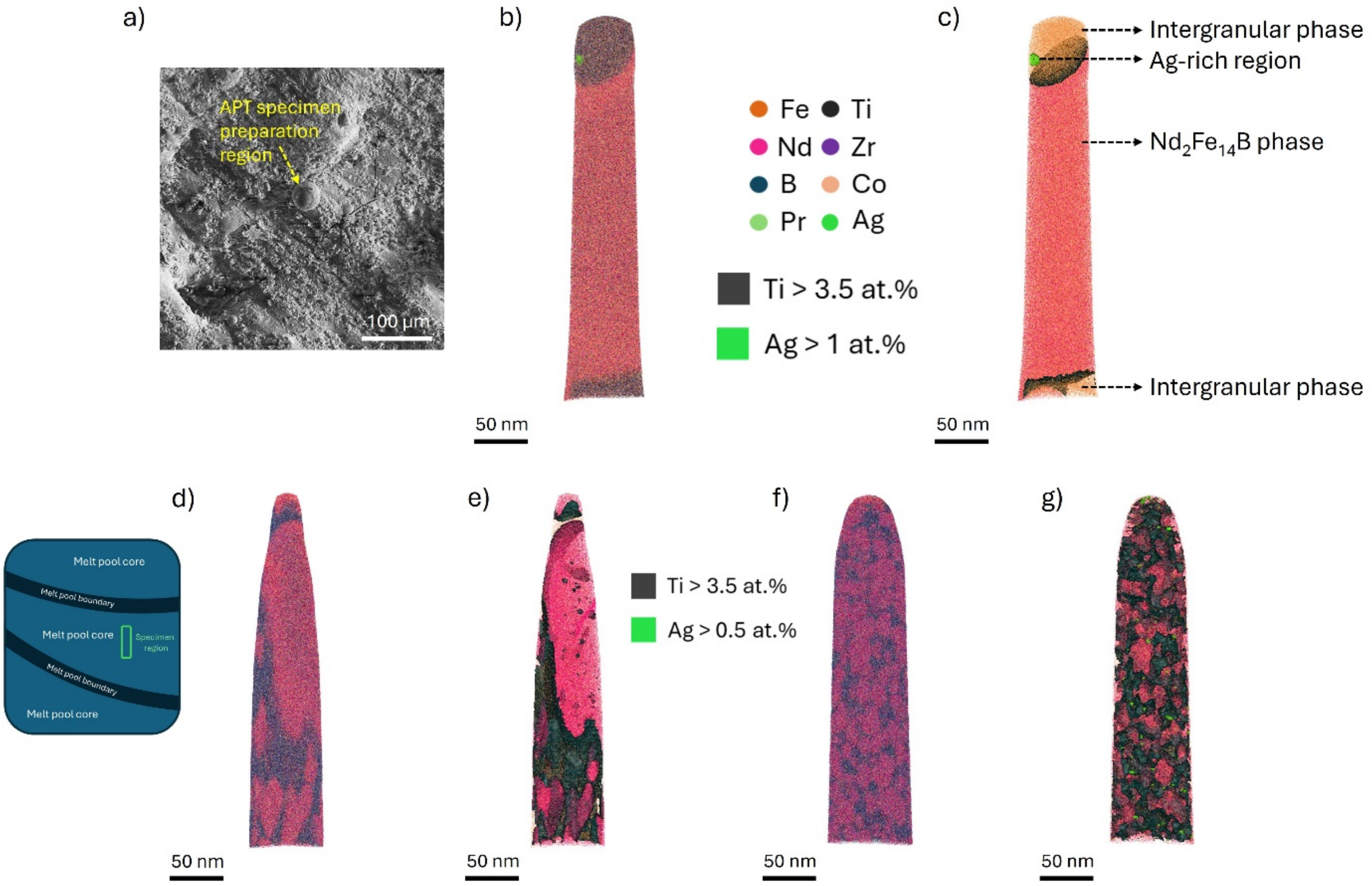

*Figure 4 a) SEM image showing a partially melted/spattered Nd-Fe-B powder particle from the feedstock found on the surface of the as-built Ag-additivated part during specimen preparation; b) 3D reconstructed APT map from the analysis of this particle; c) 3D atom map highlighting the Ti-B-Zr-rich intergranular phase with an isosurface of Ti (black) at a concentration greater than 3.5 at.%. Exemplary APT reconstructions of the (d, e) unadditivated and (f, g) Ag-additivated samples. d), f) are the 3D reconstructed atom maps; e), g) are the 3D atom maps highlighting the Ti-B-Zr-rich intergranular phase with an isosurface of Ti (grey) at a concentration greater than 3.5 at.% and Ag-rich regions with an isosurface of Ag (light green) at a concentration greater than 0.5 at.%.*

### 3.3 Distribution of Ag and other precipitate phases

In **Figure 5a**, a set of green isosurfaces with a concentration threshold of 1 at.% Ag highlights the distribution of Ag within the Ag-additivated sample. Ag-rich regions can be seen distributed throughout the specimen, adjacent to the primary magnetic phase suggesting a possible involvement in the grain refinement which is also observed in the STEM-HAADF and EDS mappings shown in **Figures 2c, d and e**. A 1D composition profile calculated along a cylindrical ROI shown in yellow in **Figure 5b**, is plotted in **Figure 5c**. The ROI crosses through the intergranular, the primary phase and the Ag-rich region. Along with approx. 3 at.% Ag and 13.51 ± 0.75 at.% Nd, these nano-sized regions in the specimen are found to be enriched with RE elements Nd and Pr, up to 15 and 3 at.%, respectively, in addition to Fe. The average elemental composition of the nano-sized Ag-rich regions after solidification as observed in APT is shown in **Table 2**.

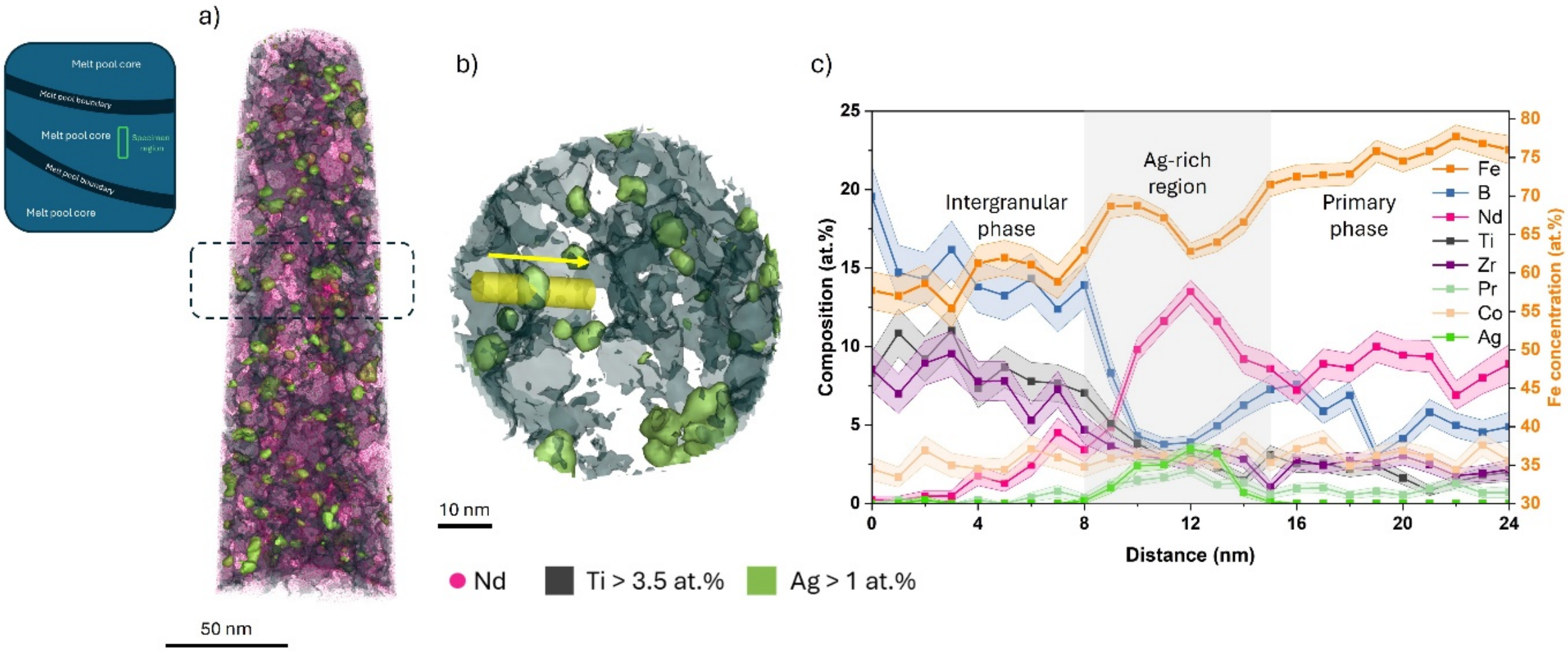

*Figure 5a) 3D atom map of the Ag-additivated sample showing the Ag distribution highlighted with an isosurface of Ag (light green) at a concentration greater than 1 at.% along with the intergranular phase regions highlighted with a Ti isosurface (grey) at a concentration greater than 3.5 at.%; b) Cross section of a selected region from the 3D reconstruction; c) 1D composition profile of the individual elements across a 5-nm-diameter cylindrical ROI from the selected region shown in (b). The concentration of Fe is plotted on the right Y-axis for better visualization.*

*Table 2 Average elemental concentrations of the nano-sized Ag-rich regions found in the Ag-additivated sample from APT.*

| Elements (at. %) | Fe | B | Nd | Ag | Ti | Zr | Pr | Co |
|---|---|---|---|---|---|---|---|---|
| | 64.3 ± 0.1 | 5.5 ± 0.05 | 15.04 ± 0.07 | 3.1 ± 0.03 | 2.6 ± 0.04 | 2.2 ± 0.03 | 2.3 ± 0.03 | 2.9 ± 0.04 |

Trace amounts of Cu, C, Cr, Al, P, O, and Si were in the remaining.

Other phases have been reported in PBF-LB/M-produced Nd-Fe-B-based magnets, with Nd-Pr-rich precipitates commonly observed [16,32]. The MQP-S feedstock powder contains alloying elements like Ti and Zr added to enhance the squareness of the hysteresis loop and promote grain refinement which can precipitate as grain boundary phases and alter the phase composition [42–46]. The presence of Zr and Ti supports the formation of the B-containing precipitate phases at the intergranular region [47] which in turn compensates for the grain boundary phases forming in the Nd-lean MQP-S feedstock powder's microstructure. Similar Ti-Zr-rich globular precipitates are found across the microstructure of the unadditivated sample. They appear randomly distributed across the intergranular phase, as shown in **Figures 6a and b** using a set of isosurfaces with a concentration threshold of 9 at.% Zr. In the Ag-additivated sample, similar Ti-Zr-rich precipitates appear, albeit with an altered morphology distributed throughout the intergranular phase region, **Figures 6c and d**. A representative 1D composition profile across a selected precipitate (**Figure 6e**) along a cylindrical ROI is shown in **Figure 6f**. The altered morphology and composition of these precipitates in the Ag-additivated sample affect the average thickness and the volume fraction of the intergranular phase as discussed in the previous section.

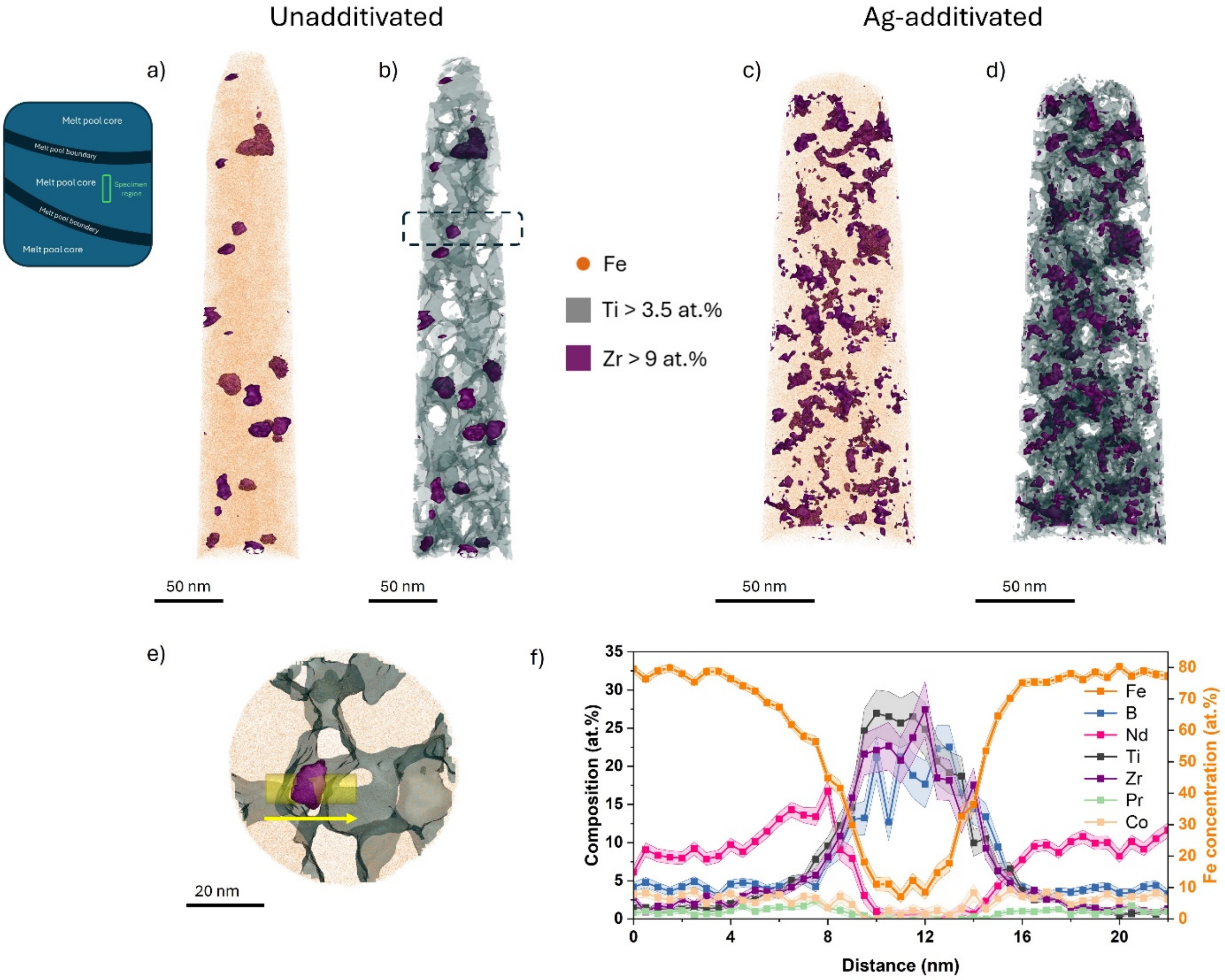

*Figure 6 3D APT reconstructions showing Ti-Zr-rich precipitate regions. a), b) and c), d) Ti-Zr-rich precipitate regions highlighted using an isosurface of Zr at a concentration greater than 9 at.% and its representation across the intergranular phase highlighted with a Ti isosurface (grey) at a concentration greater than 3.5 at.% of the unadditivated and Ag-additivated samples, respectively; e) Cross-section view of a selected Ti-Zr-B rich precipitate in the unadditivated sample; f) 1D composition profile of the individual elements across a cylindrical ROI from the selected precipitate region shown in (e). The concentration of Fe is plotted on the right Y-axis for better visualization.*

### 3.4 Proposed microstructure formation mechanism

A schematic illustration of the proposed phase evolution and microstructure formation during the solidification of the as-built PBF-LB/M processed unadditivated and Ag-additivated parts is shown in **Figure 7**. The equilibrium solidification of the stoichiometric Nd-Fe-B alloy starts with the nucleation of γ-Fe from the liquid followed by the peritectic reaction, Liquid + γ-Fe ⇌ $Nd_2Fe_{14}B$ leading to the formation of the stoichiometric 2:14:1 hard-magnetic phase [48]. Depending on the local composition (with alloying elements like Ti, Zr, Ag, Co, etc.) and the solidification conditions like cooling rate, solidification velocity, and undercooling, the phase selection can be altered [49]. Studies have shown that under conditions of sufficient undercooling (45 to 60 K), cooling rate (> $10^2$ K $s^{-1}$), and solidification velocity (> $10^{-2}$ m $s^{-1}$), direct nucleation of the $Nd_2Fe_{14}B$ phase from the melt is favored by skipping the peritectic reaction [50–52]. The PBF-LB/M processing conditions provide rapid cooling rates as high as $10^6$ K $s^{-1}$ [50] to enable the direct nucleation of the $Nd_2Fe_{14}B$ phase, thereby suppressing the soft magnetic γ-Fe leading to a nanograined microstructure in the as-built parts. During the growth of the primary $Nd_2Fe_{14}B$ phase, the excessive elements (Ti, Zr, B, etc.) were ejected into the melt, forming the surrounding intergranular phase rich in Ti, Zr, and B as observed in the unadditivated sample (**Figure 7a**).

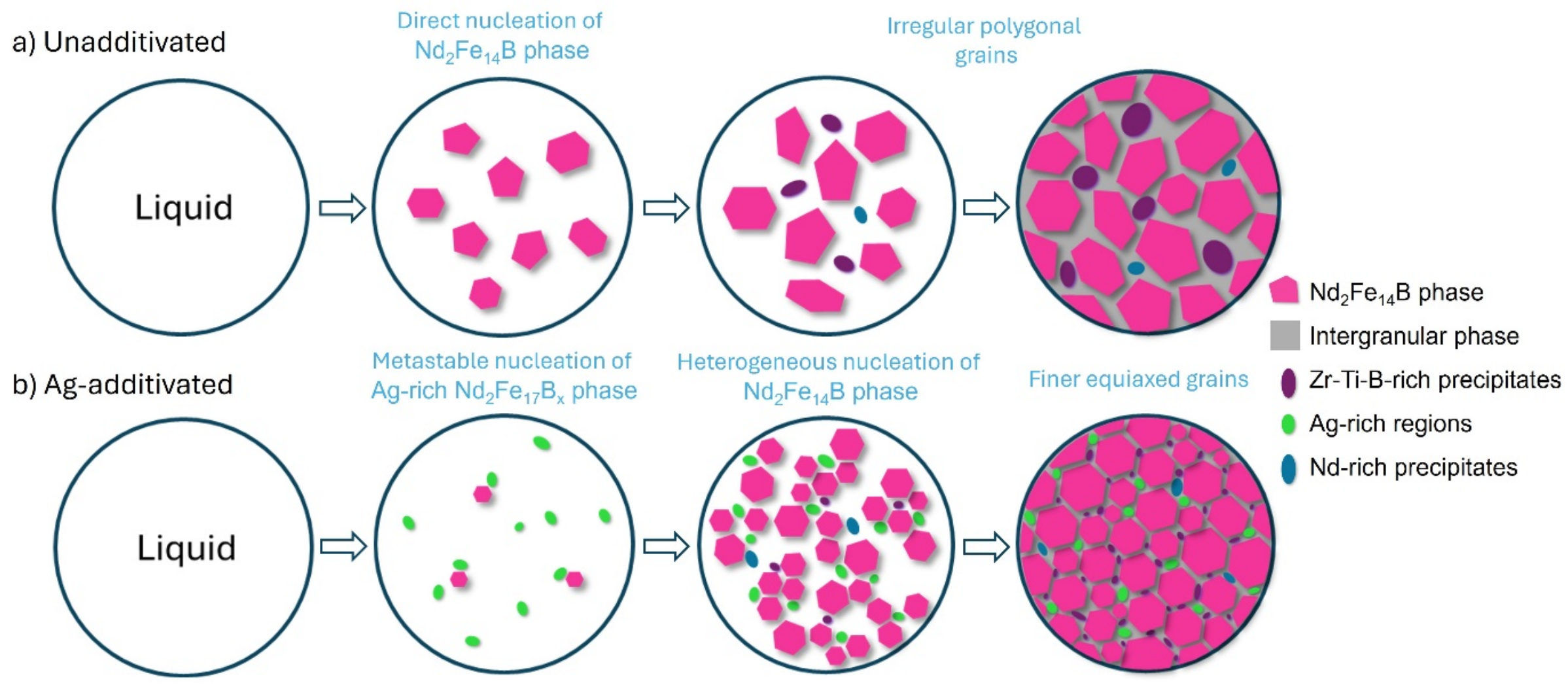

*Figure 7 Schematic illustration of the phase evolution and microstructure formation steps during the solidification of the as-built a) unadditivated and b) Ag-additivated parts.*

In the Ag-additivated sample, the microstructure formation mechanism could change with the introduction of Ag NPs on the feedstock powder surface. Since the miscibility of Ag in Fe is minimal even at higher temperatures [53], a uniform dissolution of the Ag NPs in the Nd-Fe-B system during laser melting cannot be expected. This might lead to the formation of local Ag-supersaturated regions in the liquid melt. Previous studies have reported that a metastable crystallization of the $Nd_2Fe_{17}B_x$ (x = 0–1) phase was observed in an undercooled NdFeCoZrGaB alloy similar to that of the $Sm_2Co_{17}$-type compound [54,55]. The conditions of large undercooling achieved (> 60 K), the presence of multiple alloying additions, and the kinetic advantage of forming a partially disordered structure favored the nucleation of $Nd_2Fe_{17}B_x$ over $Nd_2Fe_{14}B$ phase, which is a chemically ordered intermetallic and requires higher solute rejection [56–58]. The pseudobinary Fe–$Nd_2Fe_{14}B$ phase diagram of the ternary Nd–Fe–B system [54] also suggests a smaller thermodynamic driving force required for the metastable nucleation of $Nd_2Fe_{17}B_x$ phase compared to the $Nd_2Fe_{14}B$ phase under the conditions mentioned above. Observations of the Ag-Nd binary system (50 – 52 mol% Nd) existing in equilibrium with a Fe/$Nd_2Fe_{17}$ mixture at 1363 K have also been reported previously [59].

We propose here a similar mechanism in the case of Ag nano-additivation, where the formation of the Ag-containing $Nd_2Fe_{17}B_x$ metastable phase is promoted first from the liquid melt in the regions of Ag supersaturation caused by the melting of surface-supported Ag NPs. As the temperature of the melt decreases, the nucleation of the $Nd_2Fe_{14}B$ phase follows. The solidified metastable Ag-rich nanoscale precipitates act as heterogeneous nucleation sites for the formation of $Nd_2Fe_{14}B$ grains leading to a refined microstructure in the Ag-additivated sample (**Figure 7b**). The heterogeneous nucleation and faceted growth of the $Nd_2Fe_{14}B$ phase (**Figures 2c and d**) happens throughout the melt expelling the solutes (excess Ti, Zr, B, etc.,) to form the surrounding thin intergranular phase. It is to be noted that Ag and Fe exhibit miscibility gaps as they solidify [53,60], and the Ag-rich metastable precipitates might hence tend to reject Fe at the expense of the growth of $Nd_2Fe_{14}B$ grains. However, as the liquid cools rapidly, the Ag-rich nanophase regions are trapped in between the $Nd_2Fe_{14}B$ and intergranular phase regions, abutting the growth of the hard magnetic phase, resulting in a fine-grained microstructure. The elemental concentrations of Fe, Nd and B in the Ag-rich nanoscale precipitates obtained from APT and their relative ratios compare well with the $Nd_2Fe_{17}B_x$ (x = 0–1) metastable phase.

### 3.5 Intrinsic heat treatment effects on the microstructure

During the PBF-LB/M process, subsequent melt tracks are deposited on top of each other in the build direction which causes reheating of the underlying melt pool, often referred to as an intrinsic (or in-process) heat treatment. In addition, due to the chosen scan strategy with a 15 µm hatch spacing and a nominal laser spot size of 30 µm, individual regions within a single layer already undergo multiple thermal cycles due to overlapping laser scans. This results in an internal heat treatment effect within each layer before additional reheating occurs through subsequent layers.

The distribution, composition, and morphology of the melt pool's microstructure are altered because of this intrinsic heat treatment. Hence, it is important to substantiate that the microstructure observed in the APT is representative throughout the as-built part. The metastable Ag-rich precipitates resulting from high undercooling during solidification might get altered or not be present in the reheated or remelted regions. STEM analysis has been performed to investigate the microstructure distribution from a selected cross-section close to the center of the Ag-additivated as-built part. **Figure 8** shows the STEM-HAADF images of the selected cross-section highlighting subsequent melt tracks. The finer, equiaxed grain structure can be seen in the melt pool region surrounded by larger grains at the melt pool boundary. The average grain size of the regular equiaxed grains in the melt pool core is found to slightly increase in the built direction towards the subsequent melt pool as a result of the cooling gradient. Notably, due to the internal heat treatment within each layer, some regions of the melt pool core may already be thermally affected before the next layer is deposited.

Furthermore, STEM-EDS mappings shown in **Figures S6 and S7** taken from subsequent melt pool regions confirm the presence of Ag-rich metastable precipitates in the fine-grained melt pool core. However, at the melt pool boundary where the remelting and intrinsic heat treatment effects are significant, the microstructure is altered (**Figure S8**) with larger $Nd_2Fe_{14}B$ grains with slightly coarsened Nd-Pr-rich precipitates and less significant Ag-rich metastable phase precipitation suggesting that coalescence of precipitate phases may not only occur due to layer-wise reheating but also as a result of intra-layer heat accumulation.

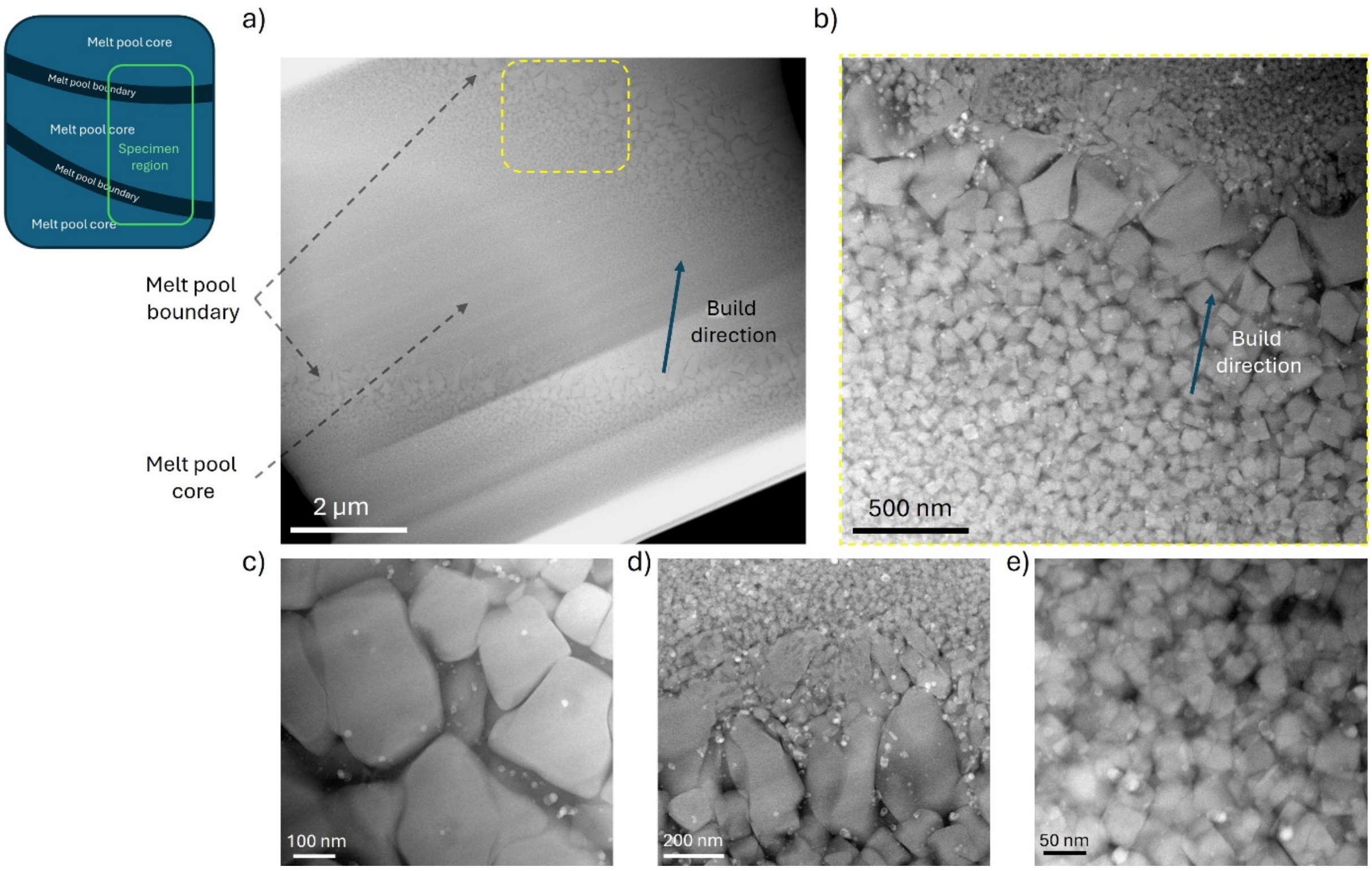

*Figure 8 STEM analysis of subsequent melt tracks. a), b) STEM-HAADF micrographs of the Ag-additivated sample highlighting subsequent melt tracks. Exemplary STEM-HAADF micrographs from a) melt pool boundary b) melt pool interface and c) melt pool core.*

Further APT analyses performed in the different regions of the melt pool in the Ag-additivated sample are shown in **Figure 9**. APT reconstructions highlighting the differences in the microstructure distribution of the melt pool boundary, melt pool interface and melt pool core similar to the STEM observations (**Figures 8c–e**) are displayed in **Figures 9a, b and c**, respectively. The set of Ti isosurfaces in grey reveals the varied grain size distributions observed in the three different regions of the melt pool. Composition profiles (i.e. proxigrams) calculated as a function of the distance to the Ti isosurfaces (Ti concentration > 3.5 at. %) of the respective 3D reconstructions highlight the distribution of individual elements between the primary $Nd_2Fe_{14}B$ phase and the intergranular phase (**Figures 9d–f**). The distribution of elements in the primary phase is similar across the different regions, whereas the concentration of elements like Fe, Ti, Zr and B in the intergranular phase differ. These elemental differences reflect not only the inter-layer heat treatment effects but also the intra-layer thermal cycling effects caused by the scan strategy and hatch spacing.

The concentration of B in the intergranular phase at the melt pool boundary amounts to around 12 at.% while in the interface region, it increases to around 15 at.% and reaches a value of 18.4 at.% at the melt pool core. The concentration of Ti and Zr amounts to around 6.7 at.% and 5.4 at.% at the melt pool boundary and increases to 8.4 at.% and 5.6 at.% at the interface region and reaches a maximum of 10.1 at.% and 6.8 at.% at the melt pool core, respectively. On the other hand, the concentration of Fe in the intergranular phase at the melt pool boundary amounts to 65.3 at.% and decreases to 59.8 at.% at the interface region and reaches a minimum of 48.1 at.% at the melt pool core making it less ferromagnetic. The elemental distribution in the intergranular phase of the melt pool core mirrors the patterns observed in **Figures 3f and h**, suggesting that metastable nucleation of Ag-rich regions during solidification favorably modifies both grain structure distribution and intergranular phase composition. Notably, the intergranular phase composition at the melt pool boundary in the Ag-additivated sample closely resembles that of the melt pool core in the unadditivated sample, while the microstructural changes induced further by Ag nano-additivation lead to enhanced magnetic properties.

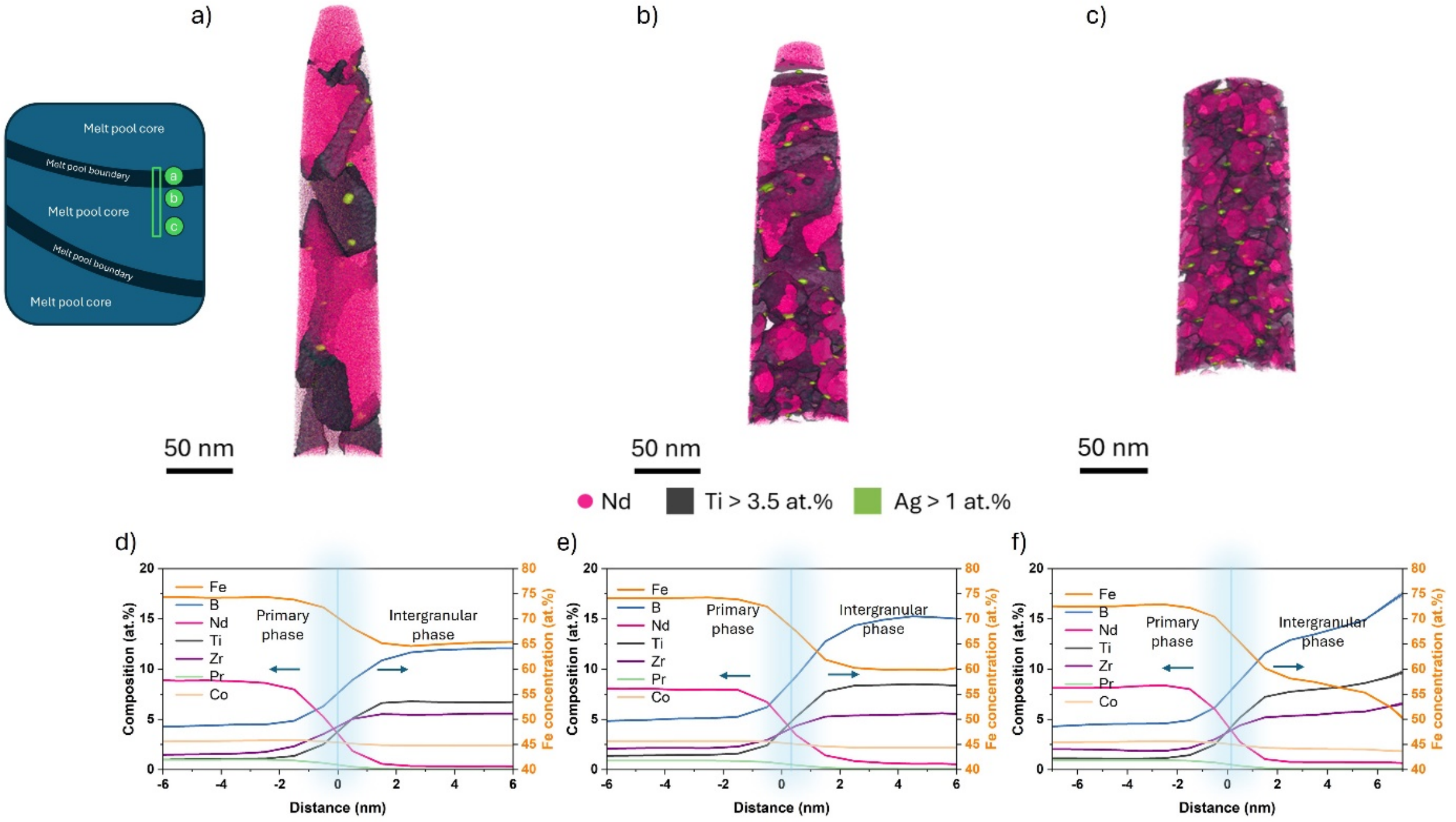

*Figure 9 3D APT reconstructions showing the microstructure and composition distribution in different regions of the melt pool in the Ag-additivated sample. a), b) and c) are the 3D atom maps from the melt pool boundary, interface close to the melt pool boundary and melt pool core, respectively, highlighting the Ti-B-Zr-rich intergranular phase with an isosurface of Ti (grey) at a concentration greater than 3.5 at.% and Ag-rich regions with an isosurface of Ag (light green) at a concentration greater than 1 at.%. d), e) and f) are the composition proxigrams constructed with Ti isosurface of concentration greater than 3.5 at.%, highlighting the distribution of individual elements between the primary and intergranular phases of the Ag-additivated sample in the melt pool boundary, interface close to the melt pool boundary and melt pool core, respectively. The concentration of Fe is plotted on the right Y-axis for better visualization.*

The morphology of the Ag-rich precipitate regions remains globular in all three regions as observed in STEM (**Figures 8c–e**) but their compositions vary. **Table 3** highlights the average elemental concentrations from APT of the selected Ag-rich precipitate regions found in the melt pool boundary, interface and the melt pool core of the Ag-additivated sample. The Nd:Ag concentration ratio varies systematically across regions: 2.29 in the melt pool boundary, 2.88 in the interface region, and 3.11 in the melt pool core. This gradient suggests that the coalescence of Ag-rich precipitates is not only driven by remelting and inter-layer heat treatment effects but also by intra-layer thermal accumulation during the PBF-LB/M process.

*Table 3 Average elemental concentrations of the selected nano-sized Ag-rich regions found in the melt pool boundary, interface and melt pool core of the Ag-additivated sample from APT.*

| Elements (at. %) | Fe | B | Nd | Ag | Ti | Zr | Pr | Co |
|---|---|---|---|---|---|---|---|---|
| Melt pool boundary | 68.3 ± 0.2 | 7.7 ± 0.08 | 7.2 ± 0.08 | 3.2 ± 0.05 | 3.6 ± 0.06 | 3.9 ± 0.07 | 1.2 ± 0.03 | 2.7 ± 0.05 |
| Interface between boundary and core | 69.01 ± 0.1 | 7.95 ± 0.06 | 7.8 ± 0.07 | 2.7 ± 0.05 | 3.6 ± 0.05 | 3.3 ± 0.04 | 1.1 ± 0.03 | 2.8 ± 0.04 |
| Melt pool core | 67.1 ± 0.1 | 8.3 ± 0.06 | 8.2 ± 0.06 | 2.7 ± 0.03 | 4.1 ± 0.04 | 3.7 ± 0.05 | 1.2 ± 0.02 | 2.6 ± 0.04 |

Trace amounts of Cu, C, Cr, Al, P, O, and Si were in the remaining.

### 3.6 Microstructure-dependent magnetic properties

To further understand the microstructure-dependent magnetic performance improvements in the as-built parts, magnetic domain structure imaging using Lorentz transmission electron microscopy (TEM) was performed [61]. **Figure 10** shows the in-focus and out-of-focus Fresnel-mode Lorentz TEM images from a selected melt pool core region in the unadditivated sample. Lorentz TEM images with a defocus of -500 µm and +500 µm are shown in **Figures 10c and d** depicting the domain walls as alternating bright and dark lines. Magnetic domain walls are found to be larger than the grain size in the melt pool core where the smaller grains coalesce to form single domains, however with individual domain walls visible around larger irregular polygonal grains (300–500 nm in diameter). Most domain walls can be seen overlapping the grain boundaries of the smaller grains where the energy of the domain boundary is lowered compared to when passing through a grain [62].

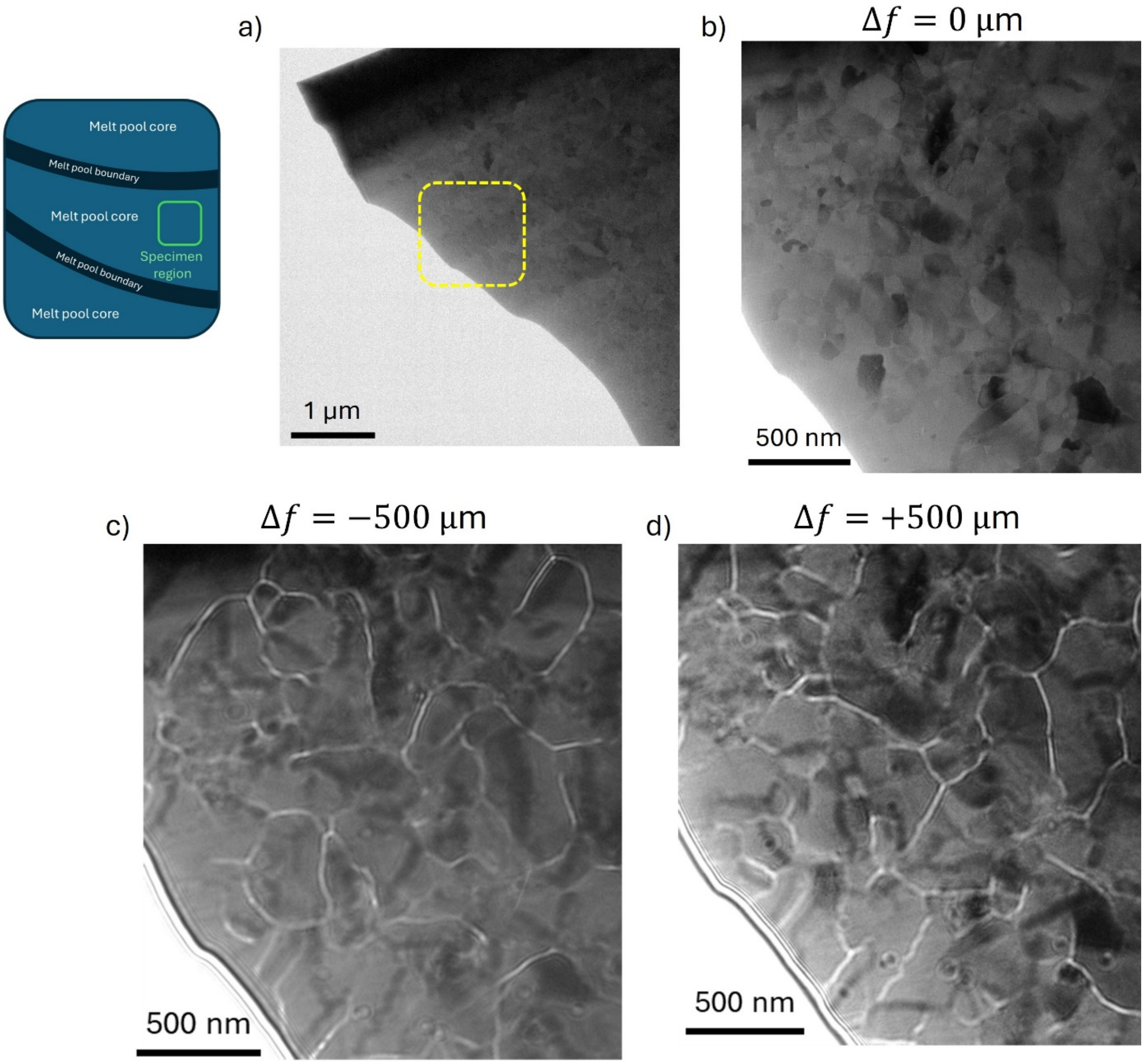

*Figure 10 Magnetic domain structure analysis of unadditivated sample using Fresnel-mode Lorentz TEM imaging. a) TEM bright field image of the specimen prepared from the unadditivated sample. Lorentz TEM images, b) in focus c) with a defocus (Δf) of -500 µm, d) with a defocus (Δf) of +500 µm showing the magnetic domain structure with alternating bright and dark lines of the region highlighted in yellow in (a).*

The Lorentz TEM images of the Ag-additivated sample shown in **Figure 11** reveal distinct variations in domain structure size across the melt pool region. Large domain structures form over coarse polygonal grains at the melt pool boundary, some covering individual grains. The presence of these larger grains at the melt pool boundary in the Ag-modified sample suggests an influence of thermal cycling on grain coarsening, and hence an influence of the build strategy, offering potentially a lever for further optimisation of the magnetic domain structure in the future.

Smaller domain structures appear toward the melt pool core, corresponding to finer equiaxed grains in the build direction, as shown in **Figures 11a and b**. The domain structures formed at the melt pool core cover several fine equiaxed grains following the grain boundaries. **Figures 11c, d and e**, showing images at 0, -500 µm and +500 µm defocus, demonstrate alternating domain wall orientations. Domain wall movement is evident near the melt pool boundary, where larger polygonal grains serve as weak pinning points.

In contrast, the melt pool core maintains finer, stable domains forming over a set of equiaxed grains, indicating stronger domain wall pinning at the less ferromagnetic Ti-Zr-B-rich intergranular phase (**Figures 9c and f**) as shown in **Figures 11 f–h**. While the unadditivated sample exhibits domain structures of 300–500 nm, the Ag-additivated sample's fine-grained equiaxed melt pool region displays significantly finer, stable domain structures. Though melt pool boundaries in the Ag-additivated sample remain weak, comparable to the unadditivated sample's domain structures, the weakest domain walls in the Ag-additivated sample match close to the strength of the strongest domains in the unadditivated sample. Despite the complex post-PBF-LB/M microstructure, characterized by multiple melt pool overlays with varying microstructural distributions, the refined microstructure achieved through Ag nano-additivation contributes to enhancing the magnetic performance.

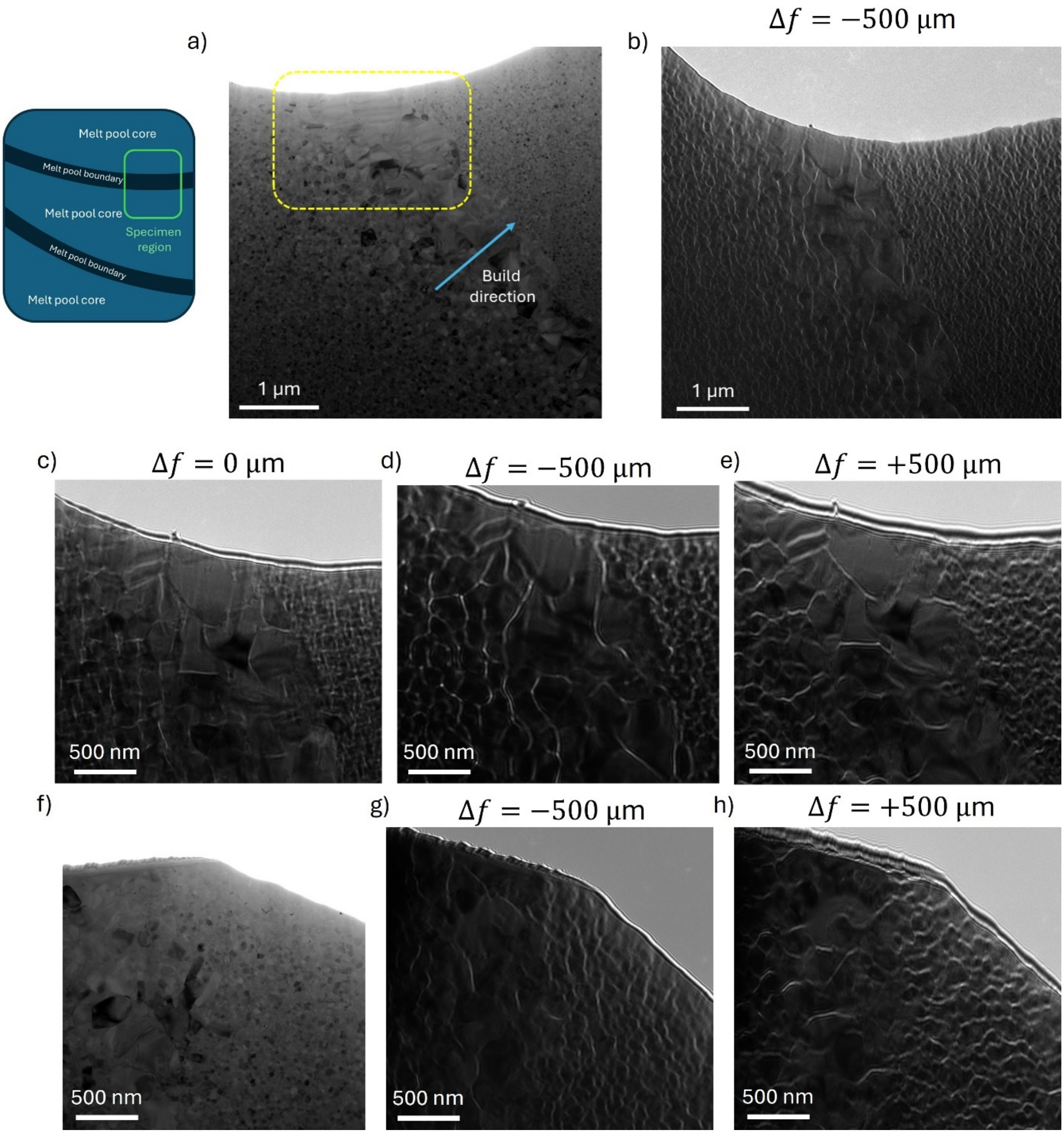

*Figure 11 Magnetic domain structure analysis of Ag-additivated sample using Fresnel-mode Lorentz TEM imaging. a) TEM bright field image of the specimen prepared from the Ag-additivated sample. b) Lorentz TEM image of the region shown in (a) with a defocus (Δf) of -500 µm. Lorentz TEM images at a higher magnification (marked in yellow in (a)), c) in focus d) with a defocus (Δf) of -500 µm, e) with a defocus (Δf) of +500 µm showing the magnetic domain walls with alternating bright and dark lines. Another representative TEM bright field image of a selected region (f) and corresponding Lorentz TEM images with a defocus (Δf) of -500 µm (g) and +500 µm (h) showing the magnetic domain walls with alternating bright and dark lines. Note: The rotation angle of the TEM bright field image is different from the Lorentz TEM images.*

## 4. Conclusions

We have demonstrated effective microstructural manipulation during near-net shape manufacturing via PBF-LB/M, establishing it as a promising method for producing Nd-Fe-B-based permanent magnets. Advanced characterization through STEM and APT revealed fundamental differences in phase composition and elemental distribution between phases resulting from Ag addition. Surface additivation of Nd-Fe-B feedstock micropowder with Ag nanoparticles yielded refined equiaxed grain structures in as-built parts through heterogeneous nucleation triggered by metastable Ag-rich nanoscale precipitates. These Ag-rich precipitate regions facilitate heterogeneous nucleation of $Nd_2Fe_{14}B$ grains, resulting in the finer distribution of Ti-Zr-rich precipitates along intergranular phase boundaries while simultaneously reducing Fe concentration. Compared to the unadditivated sample, the Ag-additivated sample exhibits a significantly finer domain structure with enhanced domain wall pinning, highlighting the impact of nano-additivation on magnetic domain behavior.

The enhanced coercivity in the Ag-additivated sample stems from three key factors: grain refinement through heterogeneous nucleation by the Ag-rich metastable phase, increased volume fraction of the primary $Nd_2Fe_{14}B$ phase, and a more amorphous, less ferromagnetic intergranular phase. This refined microstructure leads to a reduction in the domain structure and an increase in coercivity (as shown in ref [33]), as smaller grains create additional pinning sites that hinder domain wall motion. Beyond the influence of Ag nano-additivation, the selected scan strategy and thermal cycling effects during PBF-LB/M have likely contributed to the microstructural refinement. Future studies should consider isolating these effects to better understand their role in grain evolution and phase stability. Overall, nano-additivation allows for microstructure manipulation beyond what could be achieved by conventional processing and, in turn, results in enhanced magnetic properties.

## Acknowledgments

V.N. is grateful for the financial support from the International Max Planck Research School for Interface Controlled Materials for Energy Conversion (IMPRS-SurMat), now International Max Planck Research School for Sustainable Metallurgy (IMPRS-SusMet) and the Center for Nanointegration Duisburg-Essen (CENIDE). This work was financially supported by the Deutsche Forschungsgemeinschaft (DFG, German Research Foundation) within the Collaborative Research Centre / Transregio (CRC/TRR) 270, Project ID No. 405553726, subprojects A01, A10, A11, B11, and Z01. The authors thank Nick Hantke and Jan Sehrt from Ruhr University Bochum for their support with the sample production. The authors are grateful to Uwe Tezins, Christian Broß and Andreas Sturm for their support to the APT, SEM, and FIB facilities and thank Philipp Watermeyer and Volker Kree for their support to the TEM facilities at the Max Planck Institute for Sustainable Materials. V.N. greatly appreciates Aparna Saksena at the Max Planck Institute for Sustainable Materials for the help with the phase volume fraction calculations from the APT datasets.

**Declaration of competing interest**

The authors declare that they have no known competing financial interests or personal relationships that could have appeared to influence the work reported in this paper.

**CRediT authorship contribution statement**

**Varatharaja Nallathambi**: Conceptualization, Methodology, Validation, Formal analysis, Investigation, Data curation, Writing – original draft, Writing – review & editing, Visualization, Project administration. **Philipp Gabriel**: Conceptualization, Resources, Writing – original draft, Writing – review & editing. **Xinren Chen**: Investigation, Data curation, Visualization. **Ziyuan Rao**: Investigation, Data curation. **Konstantin Skokov**: Conceptualization, Writing – review & editing. **Oliver Gutfleisch**: Conceptualization, Writing – review & editing. **Stephan Barcikowski**: Conceptualization, Writing – review & editing, Supervision, Project administration, Funding acquisition. **Anna Rosa Ziefuss**: Conceptualization, Resources, Writing – original draft, Writing – review & editing, Supervision, Project administration, Funding acquisition. **Baptiste Gault**: Conceptualization, Methodology, Validation, Data curation, Writing – original draft, Writing – review & editing, Supervision, Project administration, Funding acquisition.

**Data availability statement**

The data that support the findings of this study are available from the corresponding author upon reasonable request.

# Supplementary Information

## Effect of Ag nano-additivation on microstructure formation in Nd-Fe-B magnets built by laser powder bed fusion

Varatharaja Nallathambi [a,b], Philipp Gabriel [a], Xinren Chen [b], Ziyuan Rao [b], Konstantin Skokov [c], Oliver Gutfleisch [c], Stephan Barcikowski [a], Anna Rosa Ziefuss [a]*, Baptiste Gault [b,d]*

[a] Technical Chemistry I and Center for Nanointegration Duisburg-Essen (CENIDE), University of Duisburg-Essen, 45141 Essen, Germany

[b] Max Planck Institute for Sustainable Materials, 40237 Düsseldorf, Germany

[c] Functional Materials, Institute of Material Science, Technical University of Darmstadt, 64287 Darmstadt, Germany

[d] Department of Materials, Royal School of Mines, Imperial College London, London SW72AZ, UK

* corresponding authors: b.gault@mpie.de, anna.ziefuss@uni-due.de

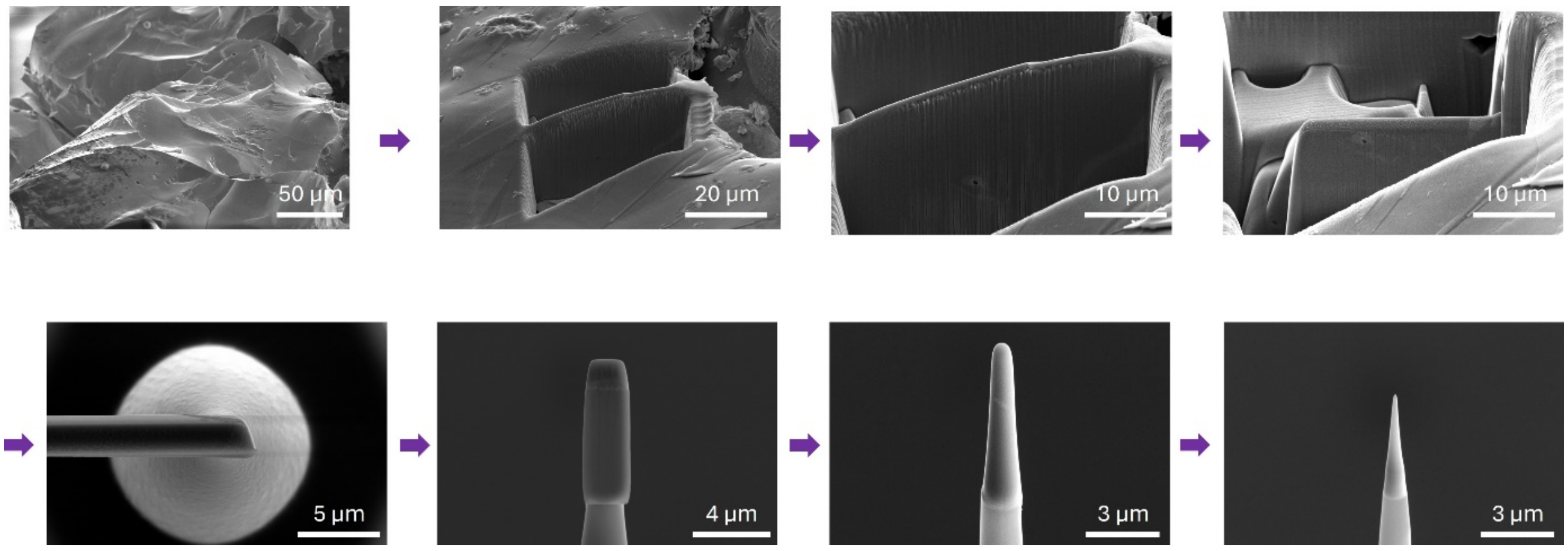

*Figure S12 APT specimen preparation steps using the standard lift-out method [1] from a PBF-LB/M as-built part.*

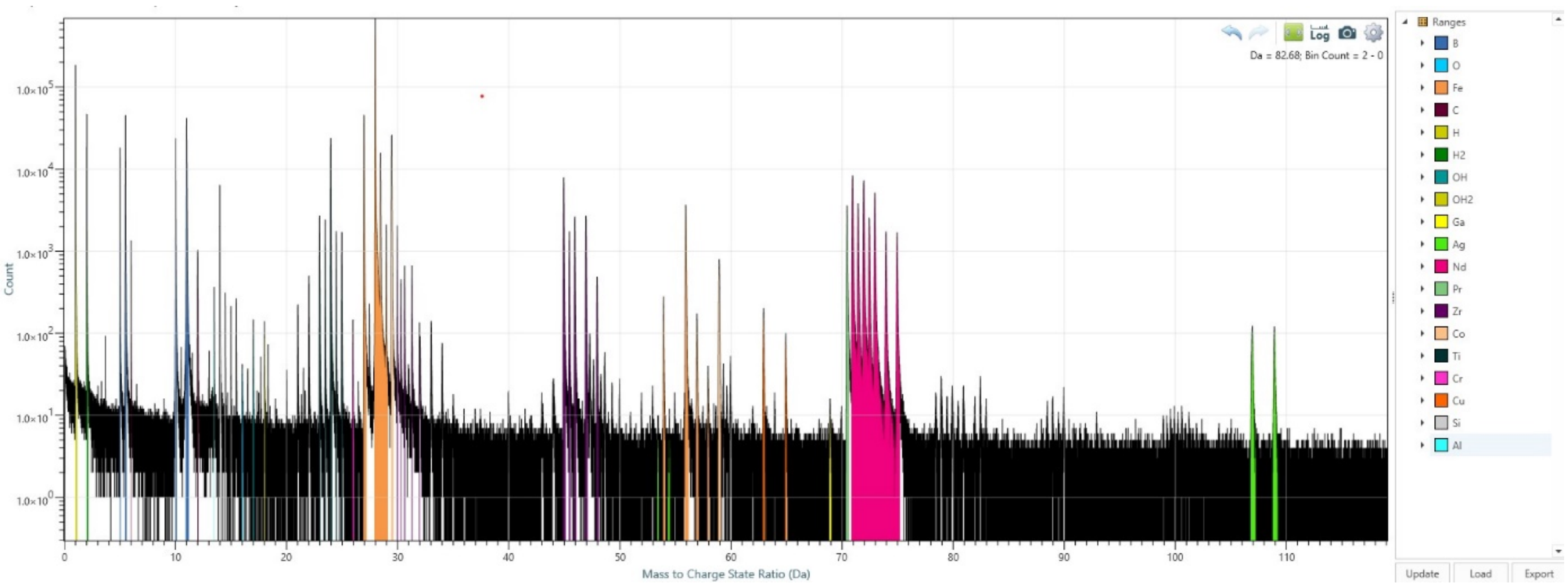

*Figure S13 Mass spectrum of the APT dataset with the respective ion ranges of the Ag-modified sample*

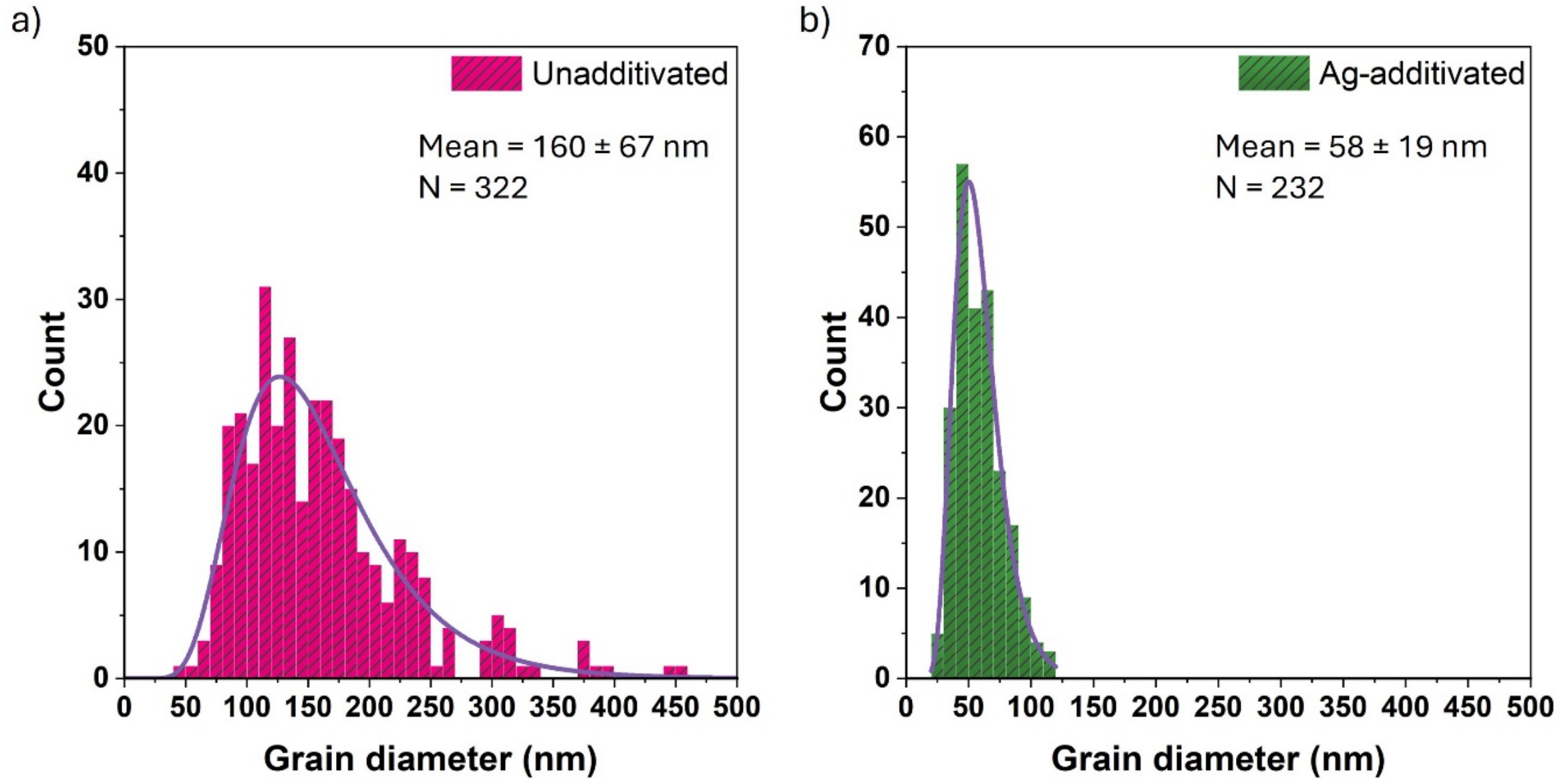

*Figure S14 Diameter equivalent grain size distribution plots with a lognormal fit for the melt pool core region of the unadditivated and Ag-additivated samples.*

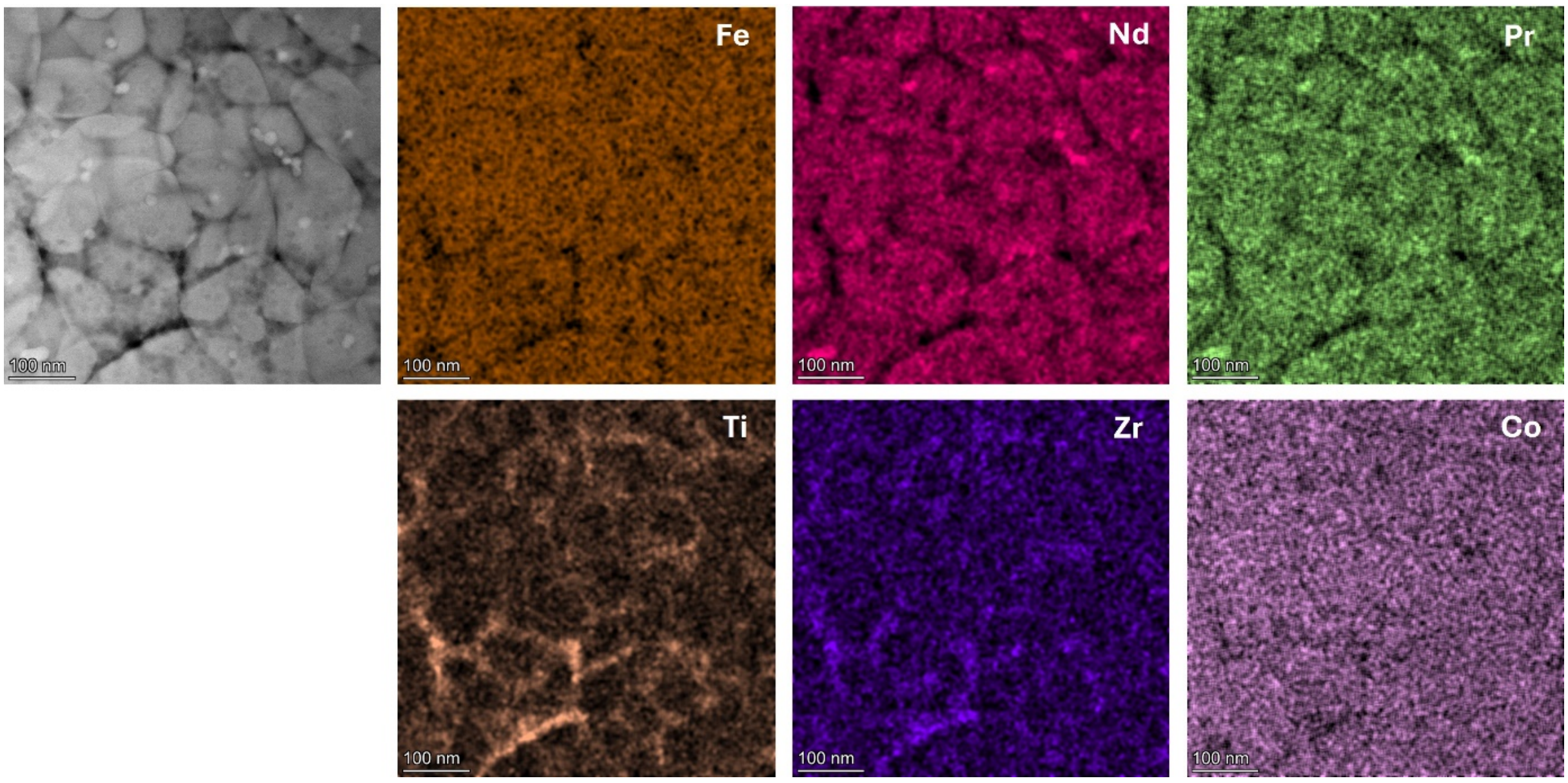

*Figure S15 STEM-EDS mappings of the melt pool core of the unadditivated sample.*

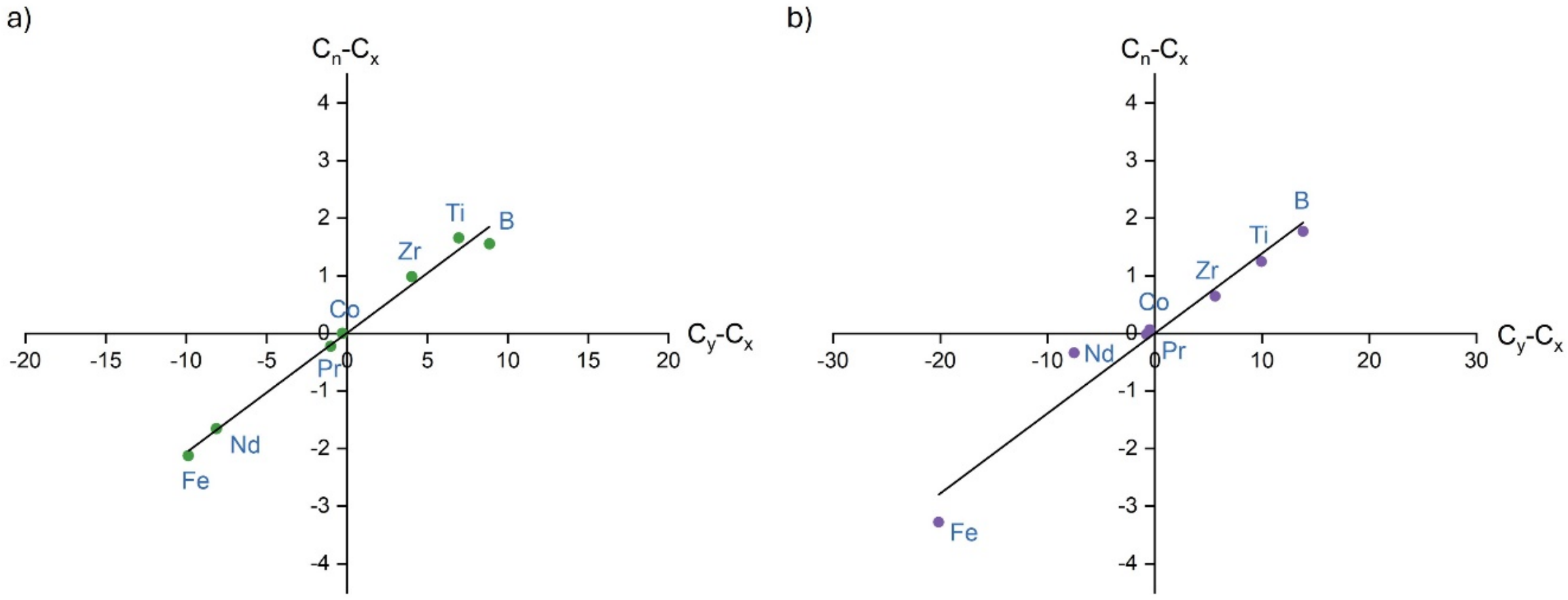

*Figure S16 Lever rule derivation plots [2,3]. The nominal concentration ($C_n$) for each element (Fe, Nd, B, etc.) minus the element content in the primary phase ($C_x$) is plotted as a function of the difference in composition between the intergranular and primary phases ($C_y$-$C_x$). The slope gives the volume fraction of the intergranular phase present in the a) unmodified and b) Ag-modified samples.*

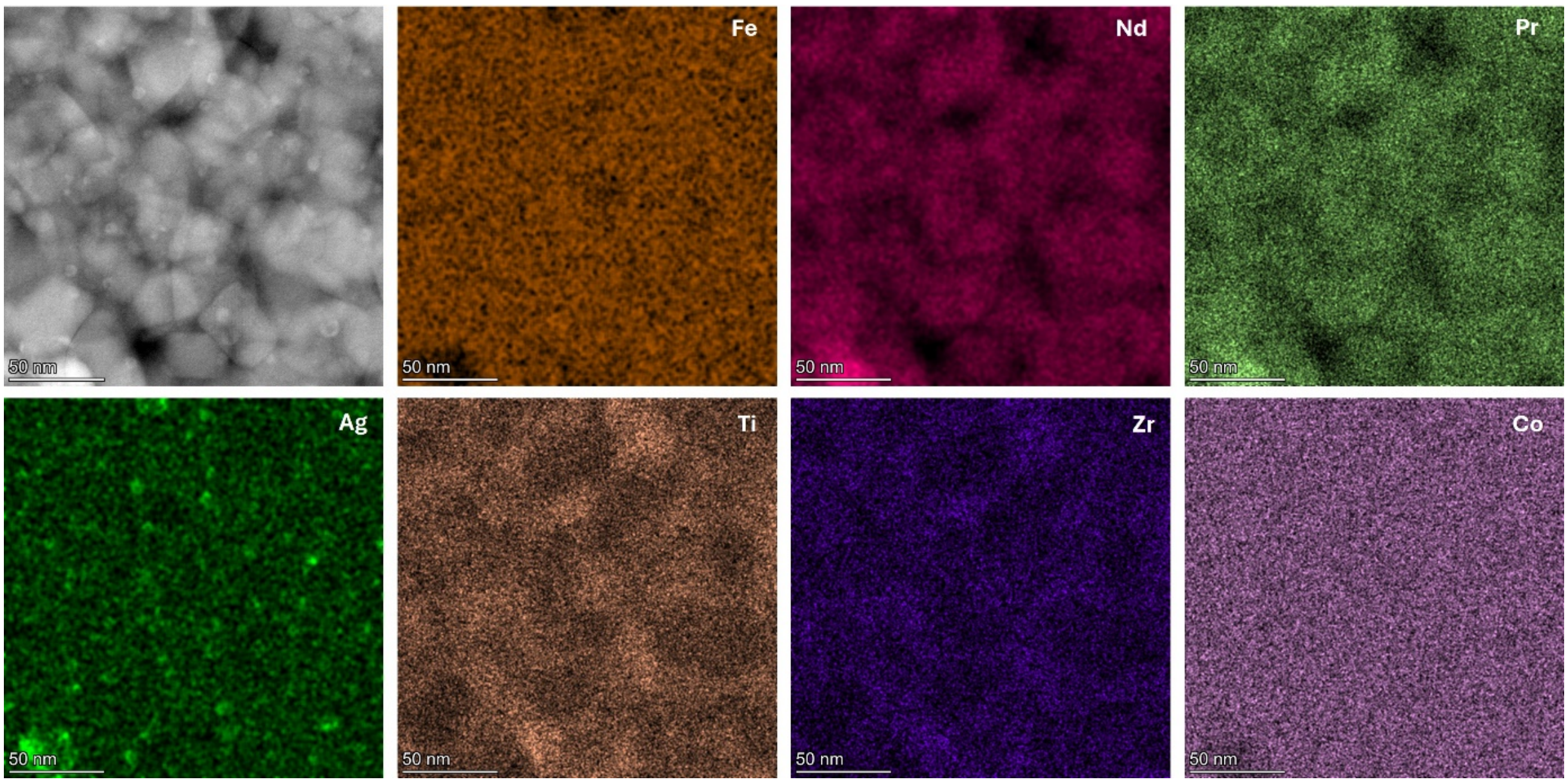

*Figure S17 STEM-EDS mappings of the Ag-additivated sample from a selected melt pool core region.*

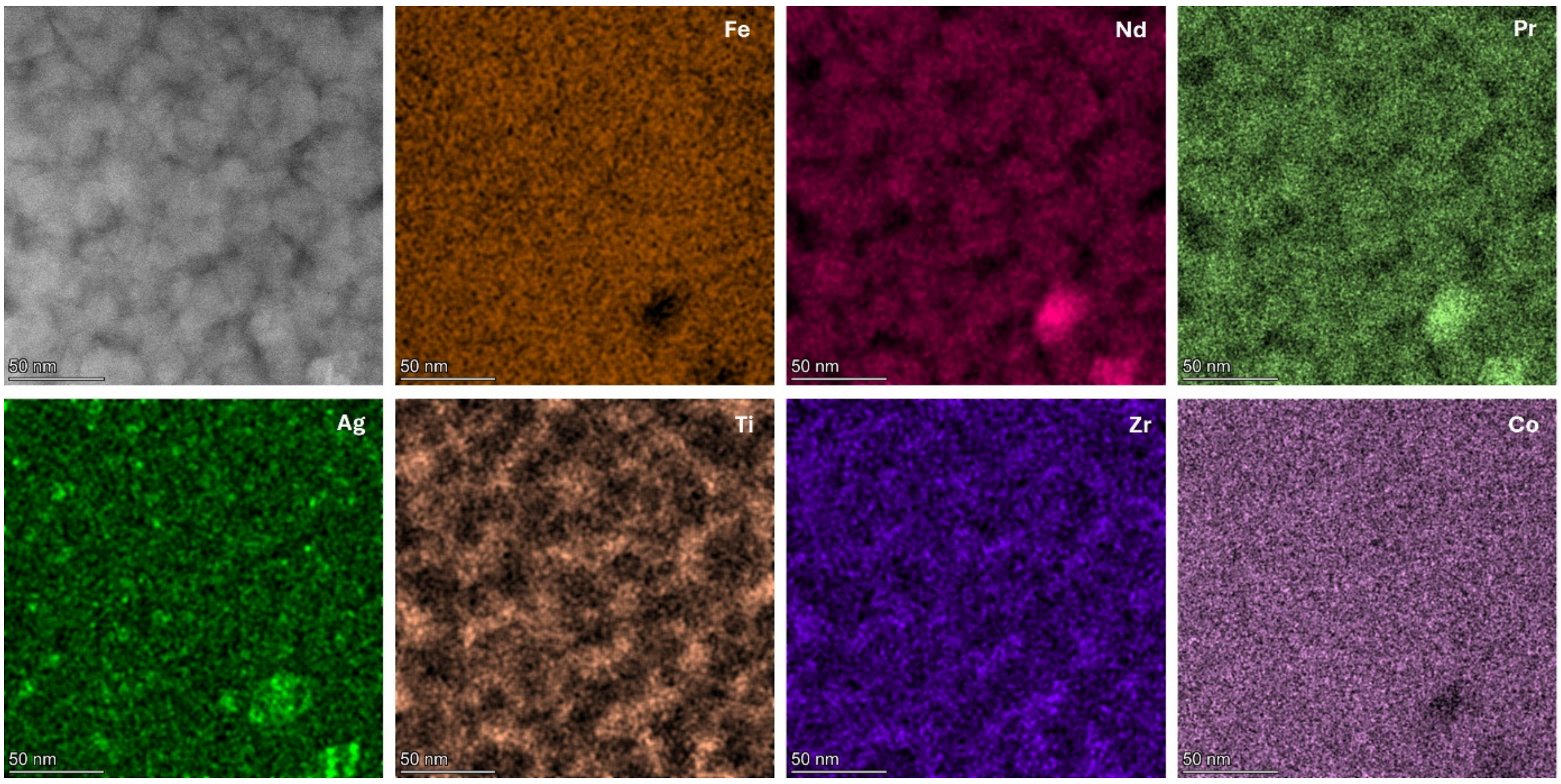

*Figure S18 STEM-EDS mappings of the Ag-additivated sample from a selected melt pool core region.*

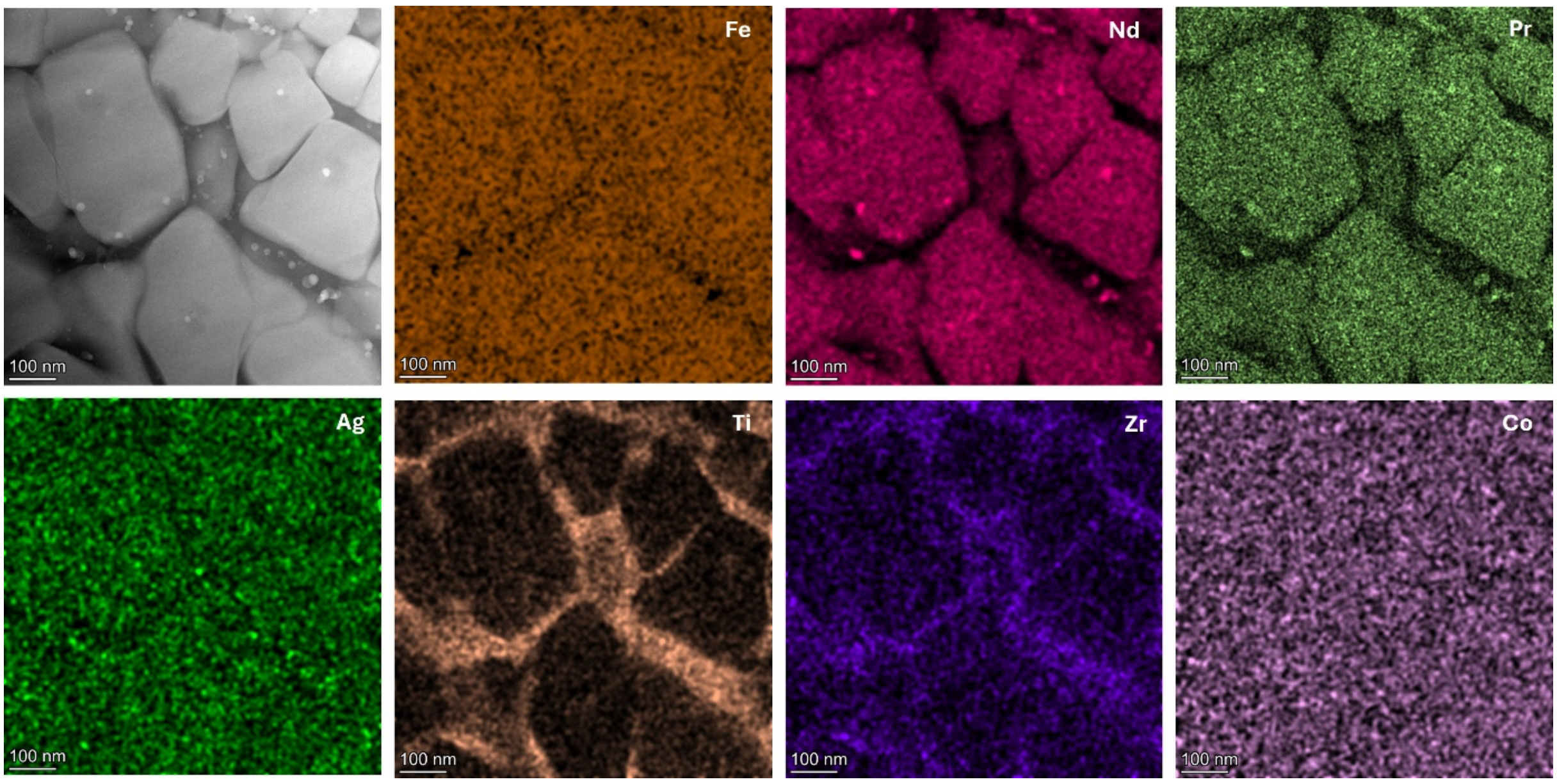

*Figure S19 STEM-EDS mappings of the Ag-additivated sample from a selected melt pool boundary region.*